\documentclass[english,journal=jctcce,manuscript=article,etalmode=truncate,maxauthors=0]{achemso}
\setkeys{acs}{etalmode=truncate,maxauthors=0}

\usepackage{amsmath}             
\usepackage{amssymb}             
\usepackage{wasysym}             
\usepackage{color}               
\usepackage{setspace}            
\usepackage{graphicx}            
\usepackage{wrapfig}             
\usepackage{siunitx}
\usepackage[dvipsnames]{xcolor}  
\usepackage{array,booktabs}      
\usepackage{mathrsfs}
\usepackage{threeparttable}
\usepackage{multirow}
\usepackage{adjustbox}

\usepackage[font=footnotesize,labelfont=bf,labelsep=period,width=0.75\textwidth]{caption}   
\usepackage[font=footnotesize,labelfont=bf,labelsep=period]{subcaption}                     
\DeclareCaptionSubType*[arabic]{figure}                                                     
\DeclareCaptionLabelFormat{subfiglabel}{Figure #2}                                          
\usepackage[capitalize]{cleveref}
\crefname{figure}{Fig.}{Figs.}
\Crefname{figure}{Figure}{Figures}
\crefname{table}{Tab.}{Tabs.}
\Crefname{table}{Table}{Tables}
\crefname{equation}{Eq.}{Eqs.}
\Crefname{equation}{Equation}{Equations}
\crefname{section}{Sec.}{Secs.}
\Crefname{section}{Section}{Sections}

\newcommand{\bra}[1]{\left\langle #1 \right\vert}         
\newcommand{\ket}[1]{\left\vert #1 \right\rangle}         

\title[Running Title]{Efficient Four-Component Dirac-Coulomb-Gaunt Hartree--Fock in Pauli Spinor Representation}
\author{Shichao Sun}
\affiliation[University of Washington]
{Department of Chemistry, University of Washington, Seattle, WA, 98195}
\alsoaffiliation[cofirst]
{Authors contributed equally}
\author{Torin F. Stetina}
\affiliation[University of Washington]
{Department of Chemistry, University of Washington, Seattle, WA, 98195}
\alsoaffiliation[cofirst]
{Authors contributed equally}
\author{Tianyuan Zhang}
\affiliation[University of Washington]
{Department of Chemistry, University of Washington, Seattle, WA, 98195}
\alsoaffiliation[cofirst]
{Authors contributed equally}
\author{Hang Hu}
\affiliation[University of Washington]
{Department of Chemistry, University of Washington, Seattle, WA, 98195}
\author{Edward F. Valeev}
\affiliation[VirginiaTech]
{Department of Chemistry, Virginia Tech, Blacksburg, VA, 24061}
\author{Qiming Sun}
\affiliation[AxiomQuant]
{AxiomQuant Investment Management LLC, Shanghai, China, 200120}
\author{Xiaosong Li}
\email{xsli@uw.edu}
\affiliation[University of Washington]
{Department of Chemistry, University of Washington, Seattle, WA, 98195}

\abbreviations{IR,NMR,UV}

\begin{document}

\begin{abstract}
Four-component Dirac Hartree--Fock is an accurate mean-field method for treating molecular systems where relativistic effects are important. However, the computational cost and complexity of the two-electron interaction makes this method less common, even though we can consider the Dirac Hartree--Fock Hamiltonian the ``ground truth" of electronic structure, barring explicit quantum-electrodynamical effects. Being able to calculate these effects is then vital to the design of lower scaling methods for accurate predictions in computational spectroscopy and properties of heavy element complexes that must include relativistic effects for even qualitative accuracy. In this work, we present a Pauli quaternion formalism of maximal component- and spin-separation for computing the Dirac-Coulomb-Gaunt Hartree--Fock ground state, with a minimal floating-point-operation count algorithm. This approach also allows one to explicitly separate different spin physics from the two-body interactions, such as spin-free, spin-orbit, and the spin-spin contributions. Additionally, we use this formalism to examine relativistic trends in the periodic table, and analyze the basis set dependence of atomic gold and gold dimer systems.

\end{abstract}

\section{Introduction}
Dirac Hartree--Fock (DHF) is a well-established method for molecules and clusters that require accurate treatment of relativistic effects such as core orbital contraction, spin-orbit coupling, and spin-spin interactions. There has been extensive research efforts utilizing  four-component DHF based on Gaussian type orbitals.\cite{Clementi91_487,Visscher94_120,Visscher97_68,Gropen97_937,Jensen99_6211,Grant98_1,Quiney00_283,Hirao01_6526,Hirao02_10122,Shiozaki13_204113} Within the DHF framework, relativistic effects can be introduced through both one-body and two-body interactions. The one-body term includes scalar relativity and one-electron spin-orbit coupling arising from the interaction with the nuclear potential. Including one-body relativistic corrections is computationally inexpensive. There are a number of widely used techniques to account for these effects in existing electronic structure methods, including one-component and two-component frameworks.

On the other hand, the two-electron operator is complicated and computationally expensive. Without going to the full quantum-electrodynamical (QED) regime, the two-electron operator includes the Dirac-Coulomb, Gaunt, and gauge (in the Coulomb gauge) terms with increasing complexity.\cite{Dyall94_2118,Dyall07_book,Wolf15_book} From the point of view of relativistic particle interactions, these terms lead to two-electron scalar relativity, spin-own-orbit, spin-other-orbit, spin-spin interactions, and also retardation effects from the finite speed of light. 
Since these important two-electron relativistic effects are computationally expensive, one often has to resort to well-designed algorithms and well-calibrated approximations to strike a balance between theoretical accuracy and computational feasibility.

Various techniques have been explored to lower the computational scaling of four-component calculations, such as the quaternion formalism,\cite{Jensen99_6211} resolution of the identity,\cite{Shiozaki13_204113,Malkin20_184101} and quasi-four component approximations.\cite{Peng06_044102,Cheng07_104106} These techniques are all based on the reduction of the dimensionality either in the matrix formalism or in the integral formation. In this work, we introduce a maximally component- and spin-separated formalism that minimizes the floating point operation (FLOP) count in the DHF formation. In the new approach introduced here, the component- and spin-separations are carried out in the restricted kinetic balance (RKB) condition within the Pauli matrix representation. All integrals are retained in the one-component electron repulsion integral (ERI) format instead of in the two-spinor or four-spinor basis.
With the Pauli components, the physics of different two-electron spin interactions is easy to recognize and categorize. This work lays the theoretical and computational foundation for practical applications of four-component Dirac Hartree--Fock and correlated relativistic electronic structure methods.

\section{Dirac-Coulomb-Gaunt Hartree--Fock Formalism in Pauli Matrix Representation}

\subsection{Brief Review of the Dirac Hartree--Fock Equation in the Four-Spinor and Two-Spinor Basis}
The one-electron Dirac equation is written in the four-spinor form $\phi = \begin{pmatrix}
\psi^L\\ \psi^S
\end{pmatrix}$, where $\psi^L$ and $\psi^S$ are two-spinors of the large and small components. The Dirac equation can be written as
\begin{equation}
    \begin{pmatrix}
    \mathbf{I}\otimes V & c \boldsymbol\sigma \cdot {\bf p} \\
    c \boldsymbol\sigma \cdot {\bf p} & \mathbf{I}\otimes (V - 2m c^2)
    \end{pmatrix}
    \begin{pmatrix}
    \psi^L \\
    \psi^S
    \end{pmatrix}
    =E 
    \begin{pmatrix}
    \psi^L \\
    \psi^S
    \end{pmatrix}
    \label{eq:1eDirac}
\end{equation}
where $\mathbf{p}=-i\boldsymbol\nabla$ is the momentum operator and $\boldsymbol\sigma$ are Pauli matrices.
\begin{align}
  \mathbf{I}&=\begin{pmatrix} 1 & 0\\ 0&1 \end{pmatrix},
\boldsymbol\sigma_x=\begin{pmatrix} 0 & 1\\ 1&0 \end{pmatrix}, \boldsymbol\sigma_y=\begin{pmatrix} 0 & -i\\ i&0 \end{pmatrix}, \boldsymbol\sigma_z=\begin{pmatrix} 1 & 0\\ 0&-1 \end{pmatrix}. \label{eq:PauliMatrix}
\end{align}

For the one-electron system, $V$ is the nuclear potential that gives rise to scalar relativistic and spin-orbit effects. For many electron systems, the leading-order two-electron interactions that enter the molecular potential are described by the Coulomb and Gaunt terms:
\begin{align}
V_{ee} &= \sum_{i=1}^N \sum_{j>i} (g^C(i,j)-g^G(i,j))\\ 
    g^C(i,j) &= \frac{1}{r_{ij}} \\
    g^G (i,j) &=\frac{\boldsymbol{\alpha}_i \cdot \boldsymbol{\alpha}_j}{r_{ij}}
\end{align}
where the components of the $\boldsymbol{\alpha}$ matrices are defined as
\begin{equation}
    {\alpha}_{i,J} = 
    \begin{pmatrix}
    {\bf 0}_2 & \boldsymbol\sigma_J \\
    \boldsymbol\sigma_J & {\bf 0}_2
    \end{pmatrix},
    \quad J = \{x, y, z\}
\end{equation}
where $i$ is $i$-th electron. Note that in this work we do not consider the gauge term which only arises in the non-quantum-electrodynamic treatment of the Coulomb gauge.

To cast the Dirac equation in a finite basis and formulate the Dirac-Coulomb Hartree--Fock (DC-HF) and Dirac-Coulomb-Gaunt Hartree--Fock (DCG-HF) Hamiltonians, the four-spinor molecular orbitals (MO) are expanded in two-spinor basis
\begin{equation}
    \psi^L_p = \sum_{{\tau}}\sum_{\mu = 1}^{N}c_{\mu\tau,p}^L {\chi}_{\mu\tau}^L, 
    \quad \psi^S_p = \sum_{\tau}\sum_{\mu = 1}^{N}c_{\mu\tau,p}^S {\chi}_{\mu\tau}^S
\end{equation}
where $\tau\in\{\alpha,\beta\}$ and $N$ is the number of spatial basis functions. The large component basis is defined as
\begin{equation}
    {\chi}_{\mu\alpha}^L = \begin{pmatrix} {\chi}_\mu \\ 0 \end{pmatrix} , \quad \quad 
    {\chi}_{\mu\beta}^L = \begin{pmatrix} 0 \\ {\chi}_{\mu} \end{pmatrix}
    \label{eq:largespinor}
\end{equation}
where ${\chi}_\mu$ denotes the spatial basis functions. 
The relationship between the large and small component two-spinor basis can be defined via the restricted kinetic balance (RKB) condition.\cite{Havriliak84_1910,Ishikawa83_111,Dyall90_25}
\begin{equation}
    {\chi}_\mu^S = \frac{1}{2mc}\boldsymbol\sigma\cdot {\bf p} {\chi}_\mu^L 
    \label{eq:rkb}
\end{equation}
This relationship ensures the correct symmetry relation (opposite parity) between the small and large components and hence the correct nonrelativistic limit of the positive energy states.\cite{Liu10_1679} For a detailed discussion of different choices of balanced bases, see Ref \citenum{Kutzelnigg11_423}. Note that the RKB condition can only be applied to the uncontracted basis functions of all the atoms in a molecule as a whole, although the idea of ``from atoms to molecule'',\cite{Cheng07_104106} where \cref{eq:rkb} is applied to the uncontracted basis functions of individual atoms, is more efficient. A brief discussion regarding the related issues will be presented in the Conclusion section.

For the Dirac-Coulomb-Gaunt Hamiltonian, the Fock matrix in the four-spinor molecular orbital representation is
\begin{align}
    f_{pq} &= h_{pq} + V_{pq}^C + V_{pq}^G\label{eq:4spinorfock}\\
    V_{pq}^C &=\sum_{i=1}^{N_e}[(pq|ii)-(pi|iq)]\label{eq:4spinorcoulomb}\\
    V_{pq}^G &= -\sum_{i=1}^{N_e}[(p\boldsymbol\alpha q|\cdot i\boldsymbol\alpha i)-(p\boldsymbol\alpha i|\cdot i\boldsymbol\alpha q)]
    \label{eq:4spinorgaunt}
\end{align}
where $N_e$ is the number of electrons, $i$ is the index for occupied orbitals, and $p,q$ are indices for general four-spinor MOs.  \Crefrange{eq:4spinorfock}{eq:4spinorgaunt} only sum over positive energy levels. Additionally, we use Mulliken notation for the two-electron integrals. We also introduce ``$|\cdot$'' notation to indicate that the integral is a dot product of $\boldsymbol\alpha$ (or $\boldsymbol\sigma$).

The expression of the Dirac-Coulomb-Gaunt Hamiltonian in the four-spinor basis is the simplest, but also the most computationally expensive to construct due to the fact that it is four times the length of the atomic orbital space. Therefore, dimension reduction techniques are usually carried out to lower the computational cost through component and/or spin separation. Unfortunately, dimension reduction procedures inevitably lead to an increased complexity in the underlying mathematical expression.

The four-component Fock matrix can be separated into four blocks ($LL, LS, SL, SS$) in the two-spinor basis. For example, the four blocks in the Dirac-Coulomb matrix are, 
\begin{align}
    V_{pq}^{C,LL} &=  \sum_{i=1}^{N_e}[(p^Lq^L|i^Li^L)+(p^Lq^L|i^Si^S)-(p^Li^L|i^Lq^L)]\label{eq:vcll}\\
    V_{pq}^{C,SS} &=  \sum_{i=1}^{N_e}[(p^Sq^S|i^Li^L)+(p^Sq^S|i^Si^S)-(p^Si^S|i^Sq^S)]\label{eq:vcss}\\
    V_{pq}^{C,LS} &=  -\sum_{i=1}^{N_e} (p^L i^L|i^S q^S)\\
    V_{pq}^{C,SL} &=  -\sum_{i=1}^{N_e} (p^S i^S|i^L q^L)
\end{align}
and the corresponding Gaunt contributions are 
\begin{align}
    V_{pq}^{G,LL} &= \sum_{i=1}^N (p^L \boldsymbol\sigma i^S|\cdot i^S \boldsymbol\sigma q^L)\\
    V_{pq}^{G,SS} &= \sum_{i=1}^N (p^S \boldsymbol\sigma i^L|\cdot i^L \boldsymbol\sigma q^S)\\
    V_{pq}^{G,LS} &= -\sum_{i=1}^N [(p^L \boldsymbol\sigma q^S|\cdot i^L \boldsymbol\sigma i^S)+(p^L \boldsymbol\sigma q^S|\cdot i^S \boldsymbol\sigma i^L)-(p^L \boldsymbol\sigma i^S|\cdot i^L \boldsymbol\sigma q^S)]\\
    V_{pq}^{G,SL} &= -\sum_{i=1}^N [(p^S \boldsymbol\sigma q^L|\cdot i^L \boldsymbol\sigma i^S)+(p^S \boldsymbol\sigma q^L|\cdot i^S \boldsymbol\sigma i^L)-(p^S \boldsymbol\sigma i^L|\cdot i^S \boldsymbol\sigma q^L)]\label{eq:vgsl}
\end{align}

The DCG Hamiltonian in the two-spinor MO basis can be transformed to the two-spinor atomic orbital basis as
\begin{align}
    V_{\mu{\tau_1},\nu{\tau_2}}^{C,LL} 
    &=\sum_{\lambda\tau_4\kappa\tau_3}D_{\lambda{\tau_4},\kappa{\tau_3}}^{LL}[ (\mu^L_{\tau_1}\nu^L_{\tau_2}|\kappa^L_{\tau_3}\lambda^L_{\tau_4})-(\mu^L_{\tau_1}\lambda^L_{\tau_4}|\kappa^L_{\tau_3}\nu^L_{\tau_2})]
    +D_{\lambda{\tau_4},\kappa{\tau_3}}^{SS} [(\mu^L_{\tau_1}\nu^L_{\tau_2}|\kappa^S_{\tau_3}\lambda^S_{\tau_4})]
    \label{eq:fLL}\\
    V_{\mu\tau_1,\nu\tau_2}^{C,SS} 
    &= \sum_{\lambda\tau_4\kappa\tau_3}D_{\lambda\tau_4,\kappa\tau_3}^{LL}[(\mu^S_{\tau_1}\nu^S_{\tau_2}|\kappa^L_{\tau_3}\lambda^L_{\tau_4})] 
    +D_{\lambda{\tau_4},\kappa{\tau_3}}^{SS}[(\mu^S_{\tau_1}\nu^S_{\tau_2}|\kappa^S_{\tau_3}\lambda^S_{\tau_4})-(\mu^S_{\tau_1}\lambda^S_{\tau_4}|\kappa^S_{\tau_3}\nu^S_{\tau_2})]
    \label{eq:fSS}\\
    V_{\mu\tau_1\nu\tau_2}^{C,LS} 
    &=-\sum_{\lambda\tau_4\kappa\tau_3}D_{\lambda\tau_4,\kappa\tau_3}^{LS}(\mu^L_{\tau_1} \lambda^L_{\tau_4}|\kappa^S_{\tau_3} \nu^S_{\tau_2})
    \label{eq:fLS}\\
    V_{\mu\tau_1\nu\tau_2}^{C,SL} 
    &=-\sum_{\lambda\tau_4\kappa\tau_3}D_{\lambda\tau_4,\kappa\tau_3}^{SL}(\mu^S_{\tau_1} \lambda^S_{\tau_4}|\kappa^L_{\tau_3} \nu^L_{\tau_2})]
    \label{eq:fSL}\\
    \notag\\
	V^{G, LL}_{\mu\tau_1,\nu\tau_2} &= \sum_{\lambda\tau_4\kappa\tau_3} D_{\lambda\tau_4,\kappa\tau_3}^{SS}  (\mu^L_{\tau_1} \boldsymbol\sigma \lambda^S_{\tau_4} | \cdot \kappa ^S_{\tau_3} \boldsymbol\sigma \nu^L_{\tau_2})
	\label{eq:fLLG}\\
	V^{G,SS}_{\mu\tau_1,\nu\tau_2} &= \sum_{\lambda\tau_4\kappa\tau_3} D_{\lambda_{\tau_4},\kappa_{\tau_3}}^{LL}  (\mu^S _{\tau_1}\boldsymbol\sigma \lambda^L _{\tau_4}| \cdot \kappa ^L_{\tau_3} \boldsymbol\sigma \nu^S_{\tau_2})
	\label{eq:fSSG}\\
	V^{G,LS}_{\mu\tau_1,\nu\tau_2}&= -\sum_{\lambda\tau_4\kappa\tau_3}  \{ D_{\lambda\tau_4,\kappa\tau_3}^{LS} (\mu^L_{\tau_1} \boldsymbol\sigma \nu^S_{\tau_2} | \cdot \kappa ^S_{\tau_3} \boldsymbol\sigma \lambda^L_{\tau_4})   
	+D_{\lambda\tau_4,\kappa\tau_3}^{SL}[  (\mu^L_{\tau_1} \boldsymbol\sigma \nu^S_{\tau_2} | \cdot \kappa ^L_{\tau_3} \boldsymbol\sigma \lambda^S_{\tau_4}) - (\mu^L _{\tau_1}\boldsymbol\sigma \lambda^S_{\tau_4} | \cdot \kappa ^L_{\tau_3} \boldsymbol\sigma \nu^S_{\tau_2})] \}
	\label{eq:fLSG}\\
	V^{G,SL}_{\mu\tau_1,\nu\tau_2} &= -\sum_{\lambda\tau_4\kappa\tau_3} \{ D_{\lambda\tau_4,\kappa\tau_3}^{SL}  (\mu^S_{\tau_1} \boldsymbol\sigma \nu^L_{\tau_2} | \cdot \kappa ^L_{\tau_3} \boldsymbol\sigma \lambda^S_{\tau_4}) 
	+D_{\lambda\tau_4,\kappa\tau_3}^{LS}[  (\mu^S_{\tau_1} \boldsymbol\sigma \nu^L_{\tau_2} | \cdot \kappa ^S _{\tau_3}\boldsymbol\sigma \lambda^L_{\tau_4}) - (\mu^S _{\tau_1}\boldsymbol\sigma \lambda^L _{\tau_4}| \cdot \kappa ^S_{\tau_3} \boldsymbol\sigma \nu^L_{\tau_2})]\}
	\label{eq:fSLG}
\end{align}
where $\tau_1,\tau_2,\tau_3,\tau_4\in\{\alpha,\beta\}$, and the density matrix in spin block formation is
\begin{equation}
    D_{\mu\tau_1,\nu\tau_2}^{XY}=\sum_{i=1}^{N_e} c_{\mu\tau_1,i}^X c_{\nu\tau_2,i}^{Y*} \quad X,Y=\{L,S\}
\end{equation}

\Crefrange{eq:fLL}{eq:fSLG} show that the Dirac-HF Hamiltonian can be conveniently constructed in the two-spinor atomic orbital basis. Based on these expressions, there are two approaches to construct the DCG-HF Hamiltonian from non-relativistic real-valued atomic basis integrals depending on how the kinetic-balance condition (\cref{eq:rkb}) is incorporated in the two-spinor basis.

In the first approach, small component bases are first generated with an appropriate kinetic-balance condition and the Dirac-HF equation can be completely reformulated in a scalar basis expansion. In this approach, only the electron-repulsion integrals are needed. However, the dimension of the small component basis is usually 2$\sim$3 times larger than that for the large component and the basis can easily become linearly dependent.

The second approach takes advantage of the restricted-kinetic-balance condition  (\cref{eq:rkb}) in the two-spinor form. \Cref{eq:rkb} gives rise to a vector form of the small component basis. As a result, the two-spinor atomic basis becomes spin-dependent and the integral-density contraction in \crefrange{eq:fLL}{eq:fSLG} require a spin-separation step. Although the construction of the Dirac-HF Hamiltonian is more complicated, the dimension of the problem remains $2N$.

Note that these two approaches are well documented and we refer the readers to Ref. \citenum{Dyall07_book} for a detailed discussion. We include these expressions herein for the sake of completeness of the theory and benchmark sections.
In this work, we introduce a further spin separation step in the Pauli matrix representation. The following method can be considered as a completely component- and spin-separated technique for the Dirac-HF equation.

\subsection{Dirac-Coulomb-Gaunt in Pauli Matrix Basis}

In this section, we will use the two-spinor integral $(\mu_{\tau_1}^L\nu_{\tau_2}^L|\kappa_{\tau_3}^S\lambda_{\tau_4}^S)$ as an example to show how component- and spin-separation can be carried out in the Pauli matrix basis.
Using the Dirac identity,\cite{Dyall07_book}
\begin{align}
(\boldsymbol{\sigma}\cdot\boldsymbol{\mu})(\boldsymbol{\sigma}\cdot\boldsymbol{\nu})&=\mathbf{I}~\left(\boldsymbol{\mu}\cdot\boldsymbol{\nu}\right)+i\boldsymbol{\sigma}\cdot\boldsymbol{\mu}\times\boldsymbol{\nu} \label{eq:diracidentity}
\end{align}
the integral can be written as
\begin{align}
   &(\mu_{\tau_1}^L\nu_{\tau_2}^L|\kappa_{\tau_3}^S\lambda_{\tau_4}^S) \notag \\
   = &
   \frac{\delta_{\tau_1 \tau_2} }{4m^2c^2}
   (\mu_{\tau_1}^L\nu_{\tau_1}^L|
   \boldsymbol\sigma(2) \cdot \mathbf{p}(2) \frac{1}{r_{12}} \boldsymbol\sigma(2) \cdot \mathbf{p}(2)  
   |\kappa_{\tau_3}^L\lambda_{\tau_4}^L) \nonumber \\
   =& \frac{\delta_{\tau_1 \tau_2}}{4m^2c^2}  [(\xi_{\tau_3}^\dagger \mathbf{I} \xi_{\tau_4})
   \boldsymbol\nabla_\kappa \cdot \boldsymbol\nabla_\lambda (\mu\nu
   |\kappa\lambda)
   + i \sum_{J=x,y,z} (\xi_{\tau_3}^\dagger \boldsymbol\sigma_J \xi_{\tau_4})
   (\boldsymbol\nabla_\kappa\times\boldsymbol\nabla_\lambda)_J
   (\mu\nu
   |\kappa\lambda)
   ]
   \label{eq:2spinprLLSS}
\end{align}
where $\boldsymbol\nabla_\kappa$ is nuclear coordinate derivative of the atomic basis $\chi_\kappa$.  $\mathbf{I}(2)$ and $\boldsymbol\sigma(2)$ are the identity and Pauli matrix for electron 2. $\xi_\tau$ is spin representation, where 
\begin{equation}
  \xi_\alpha = \begin{pmatrix} 1 \\ 0 \end{pmatrix} 
  \quad\quad 
  \xi_\beta = \begin{pmatrix} 0 \\ 1 \end{pmatrix} 
  \notag
\end{equation}
and we have 
\begin{align}
	&\xi_\alpha^\dagger \boldsymbol\sigma_x \xi_\beta = 1; \quad \xi_\beta^\dagger \boldsymbol\sigma_x \xi_\alpha = 1 \notag\\
	&\xi_\alpha^\dagger \boldsymbol\sigma_y \xi_\beta = -i; \quad \xi_\beta^\dagger \boldsymbol\sigma_y \xi_\alpha = i \notag\\
	&\xi_\alpha^\dagger \boldsymbol\sigma_z \xi_\alpha = 1; \quad \xi_\beta^\dagger \boldsymbol\sigma_z \xi_\beta = -1	\notag\\
	&\xi_\alpha^\dagger \, \mathbf{I}\, \xi_\alpha = 1; \quad \xi_\beta^\dagger \, \mathbf{I}\, \xi_\beta = 1 \notag 
\end{align}
otherwise zero. The two terms in \cref{eq:2spinprLLSS} are two-electron spin-free and spin-orbit contributions arising from the Dirac-Coulomb operator (\cref{eq:4spinorcoulomb}).

For the Gaunt term, a similar component- and spin-separation can be achieved. For example,
\begin{align}
	&(\mu^L _{\tau_1} \boldsymbol\sigma \nu^S _{\tau_2}| \cdot \kappa ^S_{\tau_3} \boldsymbol\sigma \lambda ^L _{\tau_4}) \notag \\
	=&
	\frac{1}{(2mc)^2}(\mu^L _{\tau_1} \boldsymbol\sigma(1) \boldsymbol\sigma(1)\cdot \mathbf{p}(1) \nu^L _{\tau_2} | \cdot \boldsymbol\sigma(2) \cdot \mathbf{p}(2) \kappa ^L_{\tau_3} \boldsymbol\sigma(2) \lambda ^L _{\tau_4})\nonumber \\
	=&~\xi_{\tau_1}^\dagger (1) \xi_{\tau_1} (1) \xi_{\tau_3}^\dagger (2) \xi_{\tau_4} (2) \big[\mathbf{I}(1)\mathbf{I}(2) \boldsymbol\nabla_\kappa\cdot\boldsymbol\nabla_\nu  \nonumber \\
	&+i (\mathbf{I}(1)\boldsymbol\sigma(2)+\mathbf{I}(2)\boldsymbol\sigma(1)) \cdot \boldsymbol\nabla_\kappa \times \boldsymbol\nabla_\nu \notag\\ 
	&+(\boldsymbol\sigma(1)\cdot\boldsymbol\sigma(2)) (\boldsymbol\nabla_\kappa\cdot\boldsymbol\nabla_\nu) - (\boldsymbol\sigma(1)\cdot \boldsymbol\nabla_\kappa) (\boldsymbol\sigma(2)\cdot\boldsymbol\nabla_\nu) \big]
(\mu\nu|\kappa\lambda)\label{eq:2spinorGaunt}
\end{align}
The detailed derivation can be found in the Appendix A. In  the final expression of \cref{eq:2spinorGaunt}, the first two contributions  are the two-electron spin-free and spin-orbit interactions. The final two contributions arise from the spin-spin interaction which only exists in the Gaunt term.

Careful observation of \cref{eq:2spinprLLSS} and \cref{eq:2spinorGaunt} shows that all two-spinor terms can be organized into the Pauli matrix representation of ($\mathbf{I},\boldsymbol\sigma_x,\boldsymbol\sigma_y,\boldsymbol\sigma_z$) in the form of
\begin{equation}
{\bf G}^{XY} = \frac{1}{2}{\bf G}^{XY}_s \otimes \mathbf{I} +\frac{1}{2} \sum_{J=x,y,z} {\bf G}^{XY}_J \otimes \boldsymbol\sigma_J
\end{equation}
where ${\bf G}_s$ and ${\bf G}_J$ are the spin-free and spin-dependent parts of the two-spinor general matrix/tensor, and $\{X,Y\}\in\{L,S\}$.

In fact, both two-spinor Fock $\mathbf{F}^{XY}$ and density $\mathbf{P}^{XY}$ matrices can also be written in the Pauli matrix representation,\cite{Li18_169}
\begin{align}
    {\bf Q} = \begin{pmatrix}
{\bf Q}_{\alpha\alpha} &{\bf Q}_{\beta\alpha} \\
{\bf Q}_{\beta\alpha} &{\bf Q}_{\beta\beta}
\end{pmatrix}&	=\frac{1}{2}
	\begin{pmatrix}
	{\bf Q}_s + {\bf Q}_z & {\bf Q}_x - i {\bf Q}_y\\
	{\bf Q}_x+i{\bf Q}_y  & {\bf Q}_s-{\bf Q}_z
	\end{pmatrix} =\frac{1}{2}{\bf Q}_s \otimes \mathbf{I} + \frac{1}{2}\sum_{J=x,y,z} {\bf Q}_J \otimes \boldsymbol\sigma_J  
\\
	{\bf Q}_s &= {\bf Q}_{\alpha\alpha} +{\bf Q}_{\beta\beta}\notag\\
	{\bf Q}_x &= {\bf Q}_{\alpha\beta} +{\bf Q}_{\beta\alpha}\notag\\ 
	{\bf Q}_y &= i({\bf Q}_{\alpha\beta} -{\bf Q}_{\beta\alpha})\notag\\ 
	{\bf Q}_z &= {\bf Q}_{\alpha\alpha} -{\bf Q}_{\beta\beta}\notag
\end{align}
where $\mathbf{Q}\in\{\mathbf{F}^{XY},\mathbf{P}^{XY}\}$ and $\{X,Y\}\in\{L,S\}$.

These mathematical expressions suggest that the two-spinor form of the Dirac Hamiltonian can be completely cast in the Pauli matrix representation, achieving the maximally component- and spin-separated Dirac Hamiltonian. In the next section, we show that the resulting algebraic expressions turn out to be very complicated. However, upon successful implementation, we are rewarded with an optimally efficient four-component Dirac Hartree--Fock equation, as will be shown in the Results and Discussion Section.

\subsection{Building Dirac Hartree--Fock Matrix with Pauli Components}

A unique advantage of building the DCG-HF matrix in the spin-separated Pauli spinor representation is that terms with the same integral type can be combined. For example, since the spin-free and spin-spin interactions in \cref{eq:2spinorGaunt} share a same integral type of $\boldsymbol\nabla\cdot\boldsymbol\nabla$, they can be combined in the final contraction step, resulting in a computational algorithm of potentially minimal FLOP count. Since the complete derivation is very lengthy and complex, we only present a brief derivation procedure in Appendix A and the final expressions in the main text.

\subsubsection{Dirac-Coulomb}
The non-relativistic Coulomb contribution (first and third term in \cref{eq:vcll}) only appears in the $LL$ block of the two-spinor Fock matrix, which can be written in Pauli representation as
\begin{align}
	V^{LL}_{\mu\nu,s}&=
	\sum_{\kappa\lambda } D_{\lambda\kappa, s}^{LL}[ 2(\mu\nu|\kappa\lambda)-(\mu\lambda|\kappa\nu)]\\
	V^{LL}_{\mu\nu,J}&= -\sum_{\kappa\lambda } D_{\lambda\kappa,J}^{LL}(\mu\lambda|\kappa\nu)
\end{align}
where $J=x,y,z$. Note that the non-relativistic contributions to the Pauli $J=x,y,z$ components of the $LL$ block are all exchange type. 

The rest of Dirac--Coulomb contribution involves two-center 2nd-derivatives or four-center 4th-derivatives of electron-repulsion integrals and a prefactor of $\frac{1}{(2mc)^2}$ or $\frac{1}{(2mc)^4}$.

The $LL$ block of the relativistic part (second term in \cref{eq:vcll}) of the two-spinor Dirac--Coulomb matrices in Pauli representation is
\begin{align}
	V^{C,LL}_{\mu\nu,s}&=2\sum_{\kappa\lambda}D^{SS}_{\lambda\kappa ,s}
   \boldsymbol\nabla_\kappa \cdot \boldsymbol\nabla_\lambda (\mu\nu
   |\kappa\lambda)
   + 2i \sum_{J=x,y,z} D^{SS}_{\lambda\kappa J}
   (\boldsymbol\nabla_\kappa\times\boldsymbol\nabla_\lambda)_J
   (\mu\nu
   |\kappa\lambda)
\end{align}
where $\boldsymbol\nabla_\kappa \cdot \boldsymbol\nabla_\lambda (\mu\nu|\kappa\lambda)$ and $(\boldsymbol\nabla_\kappa\times\boldsymbol\nabla_\lambda)_J(\mu\nu|\kappa\lambda)$ are the dot-product and the $J$-th component of the cross-product of the $\kappa$ and $\lambda$ center gradients.

The Dirac-Coulomb contributions to the $SS$ block (first term in \cref{eq:vcss}) are contracted with the  scalar part of the $LL$ density matrix,
\begin{align}
	V_{\mu\nu,s}^{C,SS} &= 2\sum_{\kappa\lambda} D^{LL}_{\lambda\kappa ,s} \boldsymbol\nabla_\mu\ \cdot \boldsymbol\nabla_\nu(\mu\nu|\kappa\lambda)\\
	V_{\mu\nu,J}^{C,SS} &=  2i\sum_{\kappa\lambda} D^{LL}_{\lambda\kappa ,s} (\boldsymbol\nabla_\mu\ \times \boldsymbol\nabla_\nu)_J(\mu\nu|\kappa\lambda)
\end{align}
The rest of $SS$ block, known as the $(SS|SS)$ contribution, involves $(p^Sq^S|r^Ss^S)$ type of four-center ERI 4th-derivatives. The $(SS|SS)$ Dirac-Coulomb term is often ignored in practice due to its large computational cost of ERI 4th-derivatives and its small contribution on the order of $\frac{1}{c^4}$. For the completeness of the derivation and cost analysis, the $(SS|SS)$ contribution to the Dirac-Coulomb $SS$ block is presented in Appendix B.

The Dirac-Coulomb contributions to the $LS$ block are all exchange type integrals:
\begin{align}
	V_{\mu\nu,s}^{C,LS} = \big[
	&-D_{\lambda\kappa,s}^{LS} (\boldsymbol\nabla_\kappa\cdot\boldsymbol\nabla_\nu)
	-i D_{\lambda\kappa,x}^{LS} (\boldsymbol\nabla_\kappa\times\boldsymbol\nabla_\nu)_x\notag\\
	&-iD_{\lambda\kappa,y}^{LS} (\boldsymbol\nabla_\kappa\times\boldsymbol\nabla_\nu)_y
	-iD_{\lambda\kappa,z}^{LS}  (\boldsymbol\nabla_\kappa\times\boldsymbol\nabla_\nu)_z
	\big](\mu\lambda|\kappa\nu)\\
	V_{\mu\nu,z}^{C,LS} =\big[
	&-D_{\lambda\kappa,z}^{LS} (\boldsymbol\nabla_\kappa\cdot\boldsymbol\nabla_\nu)
	- D_{\lambda\kappa,y}^{LS} (\boldsymbol\nabla_\kappa\times\boldsymbol\nabla_\nu)_x\notag\\
	&+D_{\lambda\kappa,x}^{LS} (\boldsymbol\nabla_\kappa\times\boldsymbol\nabla_\nu)_y
	-iD_{\lambda\kappa,s}^{LS}  (\boldsymbol\nabla_\kappa\times\boldsymbol\nabla_\nu)_z
	\big](\mu\lambda|\kappa\nu)\\
	V_{\mu\nu,x}^{C,LS} = \big[
	&-D_{\lambda\kappa,x}^{LS} (\boldsymbol\nabla_\kappa\cdot\boldsymbol\nabla_\nu)
	-i D_{\lambda\kappa,s}^{LS} (\boldsymbol\nabla_\kappa\times\boldsymbol\nabla_\nu)_x\notag\\
	&-D_{\lambda\kappa,z}^{LS} (\boldsymbol\nabla_\kappa\times\boldsymbol\nabla_\nu)_y
	+D_{\lambda\kappa,y}^{LS}  (\boldsymbol\nabla_\kappa\times\boldsymbol\nabla_\nu)_z
	\big](\mu\lambda|\kappa\nu)\\
	V_{\mu\nu,y}^{C,LS} = \big[
	&-D_{\lambda\kappa,y}^{LS} (\boldsymbol\nabla_\kappa\cdot\boldsymbol\nabla_\nu)
	+ D_{\lambda\kappa,z}^{LS} (\boldsymbol\nabla_\kappa\times\boldsymbol\nabla_\nu)_x\notag\\
	&-i D_{\lambda\kappa,s}^{LS} (\boldsymbol\nabla_\kappa\times\boldsymbol\nabla_\nu)_y
	- D_{\lambda\kappa,x}^{LS}  (\boldsymbol\nabla_\kappa\times\boldsymbol\nabla_\nu)_z 
	\big](\mu\lambda|\kappa\nu)
\end{align}

Additionally, for the $SL$ block, we can take advantage of the relation
\begin{equation}
	V^{C,SL} = V^{\dagger C,LS}\notag
\end{equation}
and only compute the $V^{C,LS}$ matrix elements directly.

\subsubsection{Gaunt in Pauli Components}

Next, we consider the Gaunt contribution to the Dirac Hartree--Fock matrix. The Gaunt contribution to the $LL$ and $SS$ blocks are all exchange type:
\begin{align}
	V^{G,LL}_{\mu\nu,s} =  3\big[ &D^{SS}_{\lambda \kappa,s} \boldsymbol\nabla_\lambda \cdot \boldsymbol\nabla_\kappa 
	-i D^{SS}_{\lambda \kappa,z}(\boldsymbol\nabla_\lambda \times \boldsymbol\nabla_\kappa)_z\notag\\
	-i &D^{SS}_{\lambda \kappa,x}(\boldsymbol\nabla_\lambda \times \boldsymbol\nabla_\kappa)_x  -i D^{SS}_{\lambda \kappa,y}(\boldsymbol\nabla_\lambda \times \boldsymbol\nabla_\kappa)_y  \big] (\mu\lambda|\kappa\nu)
\end{align}
\begin{align}
	V^{G,LL}_{\mu\nu,z} = & \big\{ -D^{SS}_{\lambda\kappa,z} [- (\boldsymbol\nabla_\lambda)_x (\boldsymbol\nabla_\kappa)_x  - (\boldsymbol\nabla_\lambda)_y (\boldsymbol\nabla_\kappa)_y  + (\boldsymbol\nabla_\lambda)_z (\boldsymbol\nabla_\kappa)_z] \nonumber \\
	& -i D^{SS}_{\lambda\kappa,s} (\boldsymbol\nabla_\lambda \times \boldsymbol\nabla_\kappa)_z
	-D^{SS}_{\lambda\kappa,x} [(\boldsymbol\nabla_\lambda)_x (\boldsymbol\nabla_\kappa)_z  + (\boldsymbol\nabla_\lambda)_z (\boldsymbol\nabla_\kappa)_x]\notag \\
	&-D^{SS}_{\lambda\kappa,y} [(\boldsymbol\nabla_\lambda)_y (\boldsymbol\nabla_\kappa)_z  + (\boldsymbol\nabla_\lambda)_z (\boldsymbol\nabla_\kappa)_y] \big\}(\mu\lambda|\kappa\nu)
\end{align}
\begin{align}
	V^{G,LL}_{\mu\nu,x}= & \big\{ -D^{SS}_{\lambda\kappa,x} [ (\boldsymbol\nabla_\lambda)_x (\boldsymbol\nabla_\kappa)_x  - (\boldsymbol\nabla_\lambda)_y (\boldsymbol\nabla_\kappa)_y  - (\boldsymbol\nabla_\lambda)_z (\boldsymbol\nabla_\kappa)_z] \nonumber \\
	& -i D^{SS}_{\lambda\kappa,s} (\boldsymbol\nabla_\lambda \times \boldsymbol\nabla_\kappa)_x
	-D^{SS}_{\lambda\kappa,y} [(\boldsymbol\nabla_\lambda)_x (\boldsymbol\nabla_\kappa)_y  + (\boldsymbol\nabla_\lambda)_y (\boldsymbol\nabla_\kappa)_x]\notag\\
	&-D^{SS}_{\lambda\kappa,z} [(\boldsymbol\nabla_\lambda)_x (\boldsymbol\nabla_\kappa)_z  + (\boldsymbol\nabla_\lambda)_z (\boldsymbol\nabla_\kappa)_x] \big\}(\mu\lambda|\kappa\nu)
\end{align}
\begin{align}
	V^{G,LL}_{\mu\nu,y} = & \big\{ -D^{SS}_{\lambda\kappa,y} [ -(\boldsymbol\nabla_\lambda)_x (\boldsymbol\nabla_\kappa)_x  + (\boldsymbol\nabla_\lambda)_y (\boldsymbol\nabla_\kappa)_y  - (\boldsymbol\nabla_\lambda)_z (\boldsymbol\nabla_\kappa)_z] \nonumber \\
	& -i D^{SS}_{\lambda\kappa,s} (\boldsymbol\nabla_\lambda \times \boldsymbol\nabla_\kappa)_y
	-D^{SS}_{\lambda\kappa,x} [(\boldsymbol\nabla_\lambda)_x (\boldsymbol\nabla_\kappa)_y  + (\boldsymbol\nabla_\lambda)_y (\boldsymbol\nabla_\kappa)_x]\notag\\
	&-D^{SS}_{\lambda\kappa,z} [(\boldsymbol\nabla_\lambda)_y (\boldsymbol\nabla_\kappa)_z  + (\boldsymbol\nabla_\lambda)_z (\boldsymbol\nabla_\kappa)_y] \big\}(\mu\lambda|\kappa\nu)
\end{align}

\begin{align}
	V^{G,SS}_{\mu\nu,s} &= \big[ 
	3D^{LL}_{\lambda\kappa,s}\boldsymbol\nabla_\mu\cdot \boldsymbol\nabla_\nu
	+i  D^{LL}_{\lambda\kappa,z} (\boldsymbol\nabla_\mu \times \boldsymbol\nabla_\nu)_z  +  iD^{LL}_{\lambda\kappa,x} (\boldsymbol\nabla_\mu \times \boldsymbol\nabla_\nu)_x  + iD^{LL}_{\lambda\kappa,y} (\boldsymbol\nabla_\mu \times \boldsymbol\nabla_\nu)_y\big]  (\mu\lambda|\kappa\nu) 
\end{align}
\begin{align}
	V^{G,SS}_{\mu\nu,z} = \big\{ 
	-D^{LL}_{\lambda\kappa,z}[-(\boldsymbol\nabla_\mu)_x(\boldsymbol\nabla_\nu)_x -(\boldsymbol\nabla_\mu)_y(\boldsymbol\nabla_\nu)_y + (\boldsymbol\nabla_\mu)_z(\boldsymbol\nabla_\nu)_z ]
	+3i  D^{LL}_{\lambda\kappa,s} (\boldsymbol\nabla_\mu \times \boldsymbol\nabla_\nu)_z   \nonumber \\
	-  D^{LL}_{\lambda\kappa,x} [(\boldsymbol\nabla_\mu)_x(\boldsymbol\nabla_\nu)_z + (\boldsymbol\nabla_\mu)_z(\boldsymbol\nabla_\nu)_x]
	  - D^{LL}_{\lambda\kappa,y} [(\boldsymbol\nabla_\mu)_z(\boldsymbol\nabla_\nu)_y + (\boldsymbol\nabla_\mu)_y(\boldsymbol\nabla_\nu)_z] \big\} (\mu\lambda|\kappa\nu) 
\end{align}
\begin{align}
	V^{G,SS}_{\mu\nu,x} = \big\{ 
	-D^{LL}_{\lambda\kappa,x}[(\boldsymbol\nabla_\mu)_x(\boldsymbol\nabla_\nu)_x -(\boldsymbol\nabla_\mu)_y(\boldsymbol\nabla_\nu)_y - (\boldsymbol\nabla_\mu)_z(\boldsymbol\nabla_\nu)_z ]
	+3i  D^{LL}_{\lambda\kappa,s} (\boldsymbol\nabla_\mu \times \boldsymbol\nabla_\nu)_x   \nonumber \\
	-  D^{LL}_{\lambda\kappa,z} [(\boldsymbol\nabla_\mu)_x(\boldsymbol\nabla_\nu)_z + (\boldsymbol\nabla_\mu)_z(\boldsymbol\nabla_\nu)_x]
	  - D^{LL}_{\lambda\kappa,y} [(\boldsymbol\nabla_\mu)_x(\boldsymbol\nabla_\nu)_y + (\boldsymbol\nabla_\mu)_y(\boldsymbol\nabla_\nu)_x] \big\} (\mu\lambda|\kappa\nu) 
\end{align}
\begin{align}
	V^{G,SS}_{\mu\nu,y} = \big\{ 
	-D^{LL}_{\lambda\kappa,y}[-(\boldsymbol\nabla_\mu)_x(\boldsymbol\nabla_\nu)_x +(\boldsymbol\nabla_\mu)_y(\boldsymbol\nabla_\nu)_y - (\boldsymbol\nabla_\mu)_z(\boldsymbol\nabla_\nu)_z ]
	+3i  D^{LL}_{\lambda\kappa,s} (\boldsymbol\nabla_\mu \times \boldsymbol\nabla_\nu)_y   \nonumber \\
	-  D^{LL}_{\lambda\kappa,x} [(\boldsymbol\nabla_\mu)_x(\boldsymbol\nabla_\nu)_y + (\boldsymbol\nabla_\mu)_y(\boldsymbol\nabla_\nu)_x]
	  - D^{LL}_{\lambda\kappa,z} [(\boldsymbol\nabla_\mu)_z(\boldsymbol\nabla_\nu)_y + (\boldsymbol\nabla_\mu)_y(\boldsymbol\nabla_\nu)_z] \big\} (\mu\lambda|\kappa\nu) 
\end{align}

 The Gaunt contribution to the $LS$ and $SL$ blocks includes both Coulomb and exchange integrals:
\begin{align}
	V^{G,LS}_{\mu\nu,s}&= 
	-2\sum_{\kappa\lambda} \big[ D^{LS}_{\lambda\kappa,s}\boldsymbol\nabla_\kappa\cdot\boldsymbol\nabla_\nu
	-i \sum_{J=x,y,z}D^{LS}_{\lambda\kappa,J} (\boldsymbol\nabla_\nu \times \boldsymbol\nabla_\kappa)_J\notag\\
	&- D^{SL}_{\lambda\kappa,s}\boldsymbol\nabla_\nu\cdot\boldsymbol\nabla_\lambda
	-i \sum_{J=x,y,z}D^{SL}_{\lambda\kappa,J} (\boldsymbol\nabla_\nu \times \boldsymbol\nabla_\lambda)_J
	\big] (\mu\nu|\kappa\lambda)\notag\\
	&+ \big[  D^{SL}_{\lambda\kappa ,s}\boldsymbol\nabla_\lambda\cdot \boldsymbol\nabla_\nu  
	-i D^{SL}_{\lambda\kappa,z} (\boldsymbol\nabla_\lambda \times  \boldsymbol\nabla_\nu)_z     \nonumber \\
	&-  i D^{SL}_{\lambda\kappa,y} (\boldsymbol\nabla_\lambda \times  \boldsymbol\nabla_\nu)_y   
	- iD^{SL}_{\lambda\kappa,x} (\boldsymbol\nabla_\lambda \times  \boldsymbol\nabla_\nu)_x   \big] (\mu\lambda|\kappa\nu)
\end{align}

\begin{align}
	V^{G,LS}_{\mu\nu,z} 	 &=-\big\{  
2D^{SL}_{\lambda\kappa,z}\boldsymbol\nabla_\lambda\cdot \boldsymbol\nabla_\nu 	
+2D^{SL}_{\lambda\kappa,y} (\boldsymbol\nabla_\lambda \times\boldsymbol\nabla_\nu)_x -2D^{SL}_{\lambda\kappa,x} (\boldsymbol\nabla_\lambda \times\boldsymbol\nabla_\nu)_y
-i   D^{SL}_{\lambda\kappa,s} (\boldsymbol\nabla_\lambda \times \boldsymbol\nabla_\nu)_z   \nonumber \\
	 &+D^{SL}_{\lambda\kappa,z}\big[ -(\boldsymbol\nabla_\lambda)_x(\boldsymbol\nabla_\nu)_x - (\boldsymbol\nabla_\lambda)_y(\boldsymbol\nabla_\nu)_y +(\boldsymbol\nabla_\lambda)_z(\boldsymbol\nabla_\nu)_z\big]
	   \nonumber \\
	&+  D^{SL}_{\lambda\kappa,x} \big[ (\boldsymbol\nabla_\nu)_x (\boldsymbol\nabla_\lambda)_z + (\boldsymbol\nabla_\nu)_z (\boldsymbol\nabla_\lambda)_x\big]
	+ D^{SL}_{\lambda\kappa,y} \big[ (\boldsymbol\nabla_\nu)_y (\boldsymbol\nabla_\lambda)_z + (\boldsymbol\nabla_\nu)_z (\boldsymbol\nabla_\lambda)_y\big]
	 \big\} (\mu\lambda|\kappa\nu)\notag\\
	&-2\sum_{\kappa\lambda} \big[ -i D^{LS}_{\lambda\kappa,S} (\boldsymbol\nabla_\nu \times \boldsymbol\nabla_\kappa)_z
	+D^{LS}_{\lambda\kappa,z}(\boldsymbol\nabla_\kappa \cdot \boldsymbol\nabla_\nu)
	-\sum_{K=x,y,z} D^{LS}_{\lambda\kappa,K} (\boldsymbol\nabla_\nu)_K(\boldsymbol\nabla_\kappa)_z	\notag\\
	 &+i D^{SL}_{\lambda\kappa,s} (\boldsymbol\nabla_\nu \times \boldsymbol\nabla_\lambda)_z
	+D^{SL}_{\lambda\kappa,z}(\boldsymbol\nabla_\nu \cdot \boldsymbol\nabla_\lambda)
	-\sum_{K=x,y,z} D^{SL}_{\lambda\kappa,K} (\boldsymbol\nabla_\lambda)_z(\boldsymbol\nabla_\nu)_K\big] (\mu\nu|\kappa\lambda)
\end{align}

\begin{align}
	V^{G,LS}_{\mu\nu,x}	 &=-\big\{  
2D^{SL}_{\lambda\kappa,x}\boldsymbol\nabla_\lambda\cdot \boldsymbol\nabla_\nu 	
-2D^{SL}_{\lambda\kappa,y} (\boldsymbol\nabla_\lambda \times\boldsymbol\nabla_\nu)_z +2D^{SL}_{\lambda\kappa,z} (\boldsymbol\nabla_\lambda \times\boldsymbol\nabla_\nu)_y
-i   D^{SL}_{\lambda\kappa,s} (\boldsymbol\nabla_\lambda \times \boldsymbol\nabla_\nu)_x   \nonumber \\
	 &+D^{SL}_{\lambda\kappa,x}\big[ (\boldsymbol\nabla_\lambda)_x(\boldsymbol\nabla_\nu)_x - (\boldsymbol\nabla_\lambda)_y(\boldsymbol\nabla_\nu)_y -(\boldsymbol\nabla_\lambda)_z(\boldsymbol\nabla_\nu)_z\big]
	   \nonumber \\
	&+  D^{SL}_{\lambda\kappa,y} \big[ (\boldsymbol\nabla_\nu)_x (\boldsymbol\nabla_\lambda)_y + (\boldsymbol\nabla_\nu)_y (\boldsymbol\nabla_\lambda)_x\big]
	+ D^{SL}_{\lambda\kappa,z} \big[ (\boldsymbol\nabla_\nu)_x (\boldsymbol\nabla_\lambda)_z + (\boldsymbol\nabla_\nu)_z (\boldsymbol\nabla_\lambda)_x\big]
	 \big\} (\mu\lambda|\kappa\nu)\notag\\
	&-2\sum_{\kappa\lambda} \big[ -i D^{LS}_{\lambda\kappa,s} (\boldsymbol\nabla_\nu \times \boldsymbol\nabla_\kappa)_x
	+D^{LS}_{\lambda\kappa,x}(\boldsymbol\nabla_\kappa \cdot \boldsymbol\nabla_\nu)
	-\sum_{K=x,y,z} D^{LS}_{\lambda\kappa,K} (\boldsymbol\nabla_\nu)_K(\boldsymbol\nabla_\kappa)_x	\notag\\
	 &+i D^{SL}_{\lambda\kappa,s} (\boldsymbol\nabla_\nu \times \boldsymbol\nabla_\lambda)_x
	+D^{SL}_{\lambda\kappa,x}(\boldsymbol\nabla_\nu \cdot \boldsymbol\nabla_\lambda)
	-\sum_{K=x,y,z} D^{SL}_{\lambda\kappa,K} (\boldsymbol\nabla_\lambda)_J(\boldsymbol\nabla_\nu)_K\big] (\mu\nu|\kappa\lambda)
\end{align}

\begin{align}
	V^{G,LS}_{\mu\nu,y} &=-\big\{ 
2D^{SL}_{\lambda\kappa,y}\boldsymbol\nabla_\lambda\cdot \boldsymbol\nabla_\nu 	
+2D^{SL}_{\lambda\kappa,x} (\boldsymbol\nabla_\lambda \times\boldsymbol\nabla_\nu)_z -2D^{SL}_{\lambda\kappa,z} (\boldsymbol\nabla_\lambda \times\boldsymbol\nabla_\nu)_x
-i   D^{SL}_{\lambda\kappa,s} (\boldsymbol\nabla_\lambda \times \boldsymbol\nabla_\nu)_y   \nonumber \\
	 &+D^{SL}_{\lambda\kappa,y}\big[ -(\boldsymbol\nabla_\lambda)_x(\boldsymbol\nabla_\nu)_x + (\boldsymbol\nabla_\lambda)_y(\boldsymbol\nabla_\nu)_y -(\boldsymbol\nabla_\lambda)_z(\boldsymbol\nabla_\nu)_z\big]
	   \nonumber \\
	&+  D^{SL}_{\lambda\kappa,x} \big[ (\boldsymbol\nabla_\nu)_x (\boldsymbol\nabla_\lambda)_y + (\boldsymbol\nabla_\nu)_y (\boldsymbol\nabla_\lambda)_x\big]
	+ D^{SL}_{\lambda\kappa,z} \big[ (\boldsymbol\nabla_\nu)_y (\boldsymbol\nabla_\lambda)_z + (\boldsymbol\nabla_\nu)_z (\boldsymbol\nabla_\lambda)_y\big]
	 \big\} (\mu\lambda|\kappa\nu)\notag\\
	 &-2\sum_{\kappa\lambda} \big[ -i D^{LS}_{\lambda\kappa,s} (\boldsymbol\nabla_\nu \times \boldsymbol\nabla_\kappa)_y
	+D^{LS}_{\lambda\kappa,J}(\boldsymbol\nabla_\kappa \cdot \boldsymbol\nabla_\nu)
	-\sum_{K=x,y,z} D^{LS}_{\lambda\kappa,K} (\boldsymbol\nabla_\nu)_K(\boldsymbol\nabla_\kappa)_y	\notag\\
	 &+i D^{SL}_{\lambda\kappa,s} (\boldsymbol\nabla_\nu \times \boldsymbol\nabla_\lambda)_y
	+D^{SL}_{\lambda\kappa,y}(\boldsymbol\nabla_\nu \cdot \boldsymbol\nabla_\lambda)
	-\sum_{K=x,y,z} D^{SL}_{\lambda\kappa,K} (\boldsymbol\nabla_\lambda)_y(\boldsymbol\nabla_\nu)_K\big] (\mu\nu|\kappa\lambda)
\end{align}

Additionally, for the $SL$ block, we have the symmetry
\begin{equation}
	V^{G,SL} = V^{\dagger G,LS}\notag
\end{equation}

\subsubsection{Core Hamiltonian in Pauli Components}
Finally, the core Hamiltonian is then defined as
\begin{align}
h^{LL}_{\mu\nu,s} &= 2\langle\mu|V|\nu\rangle,\quad h^{LL}_{\mu\nu,J}=0\\
h^{LS}_{\mu\nu,s} &= 2\langle\mu|T|\nu\rangle,\quad h^{LS}_{\mu\nu,J}=0\\
h^{SS}_{\mu\nu,s} &= 2\bigg[\frac{1}{4m^2c^2}\boldsymbol\nabla_\mu\cdot\boldsymbol\nabla_\nu\langle\mu|V|\nu\rangle-\langle\mu|T|\nu\rangle\bigg]\\
h^{SS}_{\mu\nu,J}&=\frac{2i}{4m^2c^2}(\boldsymbol\nabla_\mu \times \boldsymbol\nabla_\nu)_J \langle\mu|V|\nu\rangle
\end{align}
where $J=x,y,z$. $T$ and $V$ are kinetic energy and electron-nuclear potential energy operators.

\subsection{Dirac Hartree--Fock Equation}

Collecting all Pauli components for core-Hamiltonian, Dirac-Coulomb, and Gaunt terms, the Fock matrix in Pauli representation can be constructed:
\begin{align}
    \mathbf{F}^{XY}_\Gamma = \mathbf{h}^{XY}_\Gamma + \mathbf{V}^{XY}_\Gamma + \frac{1}{(2mc)^2}\mathbf{V}^{C,XY}_\Gamma + \frac{1}{(2mc)^2}\mathbf{V}^{G,XY}_\Gamma + \frac{1}{(2mc)^4}\mathbf{V}^{C(4), XY}_{\Gamma}
\end{align}
where $X,Y \in\{L,S\}$ and $\Gamma\in\{s,x,y,z\}$. The corresponding two-spinor Fock matrix and density matrix can be constructed from their Pauli components:
\begin{equation}
{\bf F}^{XY} =\frac{1}{2}
\begin{pmatrix}
{\bf F}^{XY}_s+{\bf F}^{XY}_z &   {\bf F}^{XY}_x - i {\bf F}^{XY}_y\\
{\bf F}^{XY}_x+i{\bf F}^{XY}_y  & {\bf F}^{XY}_s-{\bf F}^{XY}_z
\end{pmatrix}
,\quad 
{\bf P}^{XY} =\frac{1}{2}
\begin{pmatrix}
{\bf P}^{XY}_s+{\bf P}^{XY}_z &   {\bf P}^{XY}_x - i {\bf P}^{XY}_y\\
{\bf P}^{XY}_x+i{\bf P}^{XY}_y  & {\bf P}^{XY}_s-{\bf P}^{XY}_z
\end{pmatrix}
\end{equation}

Finally, the four-component Dirac equation in the restricted-kinetic-balanced condition is expressed as
\begin{align}
     \begin{pmatrix}
         {\bf F}^{LL} & {\bf F}^{LS} \\
         {\bf F}^{SL} & {\bf F}^{SS}
     \end{pmatrix}
     \begin{pmatrix}
         {\bf C}^+_L &{\bf C}^-_L  \\
         {\bf C}^+_S &{\bf C}^-_S
     \end{pmatrix}
     =
     \begin{pmatrix}
         \mathbf{I}\otimes{\bf S} & {\bf 0} \\
         {\bf 0} & \mathbf{I} \otimes \frac{1}{2mc^2}{\bf T}
     \end{pmatrix}
     \begin{pmatrix}
         {\bf C}^+_L &{\bf C}^-_L  \\
         {\bf C}^+_S &{\bf C}^-_S
     \end{pmatrix}
     \begin{pmatrix}
         \boldsymbol{\epsilon}^+ &{\bf 0}  \\
         {\bf 0} &\boldsymbol{\epsilon}^-
     \end{pmatrix}
     \label{eq:fourcomp}
\end{align}
where $\mathbf{S}$ and $\mathbf{T}$ are the one-component overlap and kinetic matrices. $  \boldsymbol{\epsilon}^+ $, $ \boldsymbol{\epsilon}^-$ are the eigenvalues with corresponding molecular orbitals $\begin{pmatrix} {\bf C}^+_{L} \\ {\bf C}^+_{S} \end{pmatrix}$ and $\begin{pmatrix} {\bf C}^-_{L} \\ {\bf C}^-_{S} \end{pmatrix}$ for positive-  and  negative-energy solutions, respectively.

\vspace{20pt}
Although the resulting integral-density contraction scheme during the four-component Fock-build is complicated, there are many unique advantages that lead to a highly efficient four-component method:
\begin{itemize}
    \item The algorithm presented here achieves the maximal component- and spin-separation so that all contractions are essentially done in the one-component framework. In such a framework, terms that use the same type of electron-repulsion integrals can be effectively combined, possibly leading to a minimal FLOP-count algorithm.
    \item The Pauli matrix representation is ideal for including external electromagnetic field perturbations, because the effects of each component of the electromagnetic field are  completely decoupled and spin/component-separated.\cite{Li20_4533,Li19_6824,Li19_739,Li19_3162,Li19_348} 
    \item We also argue that the spin-separation in Pauli matrix representation allows for a clear understanding of how different types of spin physics, including two-electron spin-free, spin-orbit, and spin-spin interactions are manifested in electronic structure and light-matter coupling.
\end{itemize}
In the next section, we substantiate these claims with more details of the computational implementation, and an analysis of computational cost.

\section{Benchmark and Discussion}

\subsection{Analysis of Computational Cost}

As for any form of four-component DHF implementation, the assembly of integrals needed for the Pauli matrix formalism presented in this work also involves computations of two-center 2nd-derivatives of electron repulsion integrals (ERIs). If the $(SS|SS)$ Dirac-Coulomb term is requested, additional four-center 4th-derivatives of ERIs are needed. 
Although there is no need to assemble real-valued ERIs into two-spinor or four-spinor forms, it is necessary to construct unique integrals needed throughout the Dirac-Coulomb-Gaunt Fock build. For example, in addition to the normal ERI terms, $(\mu\nu|\kappa\lambda)$, four integrals, including $\boldsymbol\nabla\cdot\boldsymbol\nabla$, $(\boldsymbol\nabla\times\boldsymbol\nabla)_x$, $(\boldsymbol\nabla\times\boldsymbol\nabla)_y$, and $(\boldsymbol\nabla\times\boldsymbol\nabla)_z$, are constructed from two-center ERI 2nd-derivatives in order to form the Dirac-Coulomb Fock matrix with an algorithm using an optimal FLOP count. In the current development, a total of 24 unique integrals are constructed from ERI and ERI 2nd-derivatives in order to form the Dirac-Coulomb-Gaunt Fock matrix without the $(SS|SS)$ contributions.

The following discussion and cost analyses are based on the in-core formalism in which all unique integrals are computed and stored in memory. The computational cost for the $(SS|SS)$ contribution is separated from the Dirac-Coulomb term as it is often ignored in practical calculations. The four-spinor form of the Dirac-Fock build is not considered here because it is mathematically more computationally expensive than those in the two-spinor, scalar basis, or Pauli matrix representation. In the case of the scalar basis formalism, we assume the small component basis is generated through the action of the momenta $\mathbf{p}$ or the unrestricted-kinetic-balance (UKB) condition. 
For the UKB scalar-basis based method, the following analysis assumes an algorithm of optimal FLOP and storage with all integrals in the scalar one-component form. A detailed analysis for the UKB scalar basis is presented in Appendix C.

\begin{table*}[h]
\centering
\begin{threeparttable}
\caption{Analysis of memory requirement in terms of the equivalent number of unique real-valued one-component 4-index electron-repulsion-integral tensors needed for forming the four-component Dirac-Coulomb-Gaunt Hamiltonian. $N$ number of atomic basis functions are used in this estimate. Restricted-kinetic-balanced (RKB) basis is compared with the unrestricted-kinetic-balanced (UKB) scalar basis. Dense matrices with no integral or matrix symmetry are considered.}
\label{tab:RAM}
\footnotesize

\begin{tabular*}{6.5in}{@{\extracolsep{\fill} } lcccc}
\hline

\hline
\multirow{2}{*}{Four-Component Scheme} & \multicolumn{3}{c}{Equivalent \# of Real-Valued $N^4$ ERI Tensors} & \multirow{2}{*}{Total}\\
\cline{2-4}
 & 
 Dirac-Coulomb$^{[a]}$  & $(SS|SS)$& Gaunt  &  \\ 
\hline

Pauli Matrix RKB Basis &  5& 31 & 19 & 55\\
Two-Spinor RKB Basis$^{[b]}$ &  64 & 32& 64 & 160 \\
UKB Scalar Basis$^{[c]}$ & 7.25 & 39 & 6.25 &52.5  \\
\hline
\hline
\end{tabular*}
\begin{tablenotes}
\item[a] Dirac-Coulomb without the $(SS|SS)$ contribution.
\item[b] Two-spinor integrals are stored as complex-valued quantities, occupying two double-precision storage.
\item[c] For scalar-basis based approach, we assume there are $\frac{N}{2}$ number of $L>0$ atomic basis function in this set, resulting in a total of 2.5$N$ number of smaller component basis functions. For the memory requirement estimation, see Appendix C.
\end{tablenotes}
\end{threeparttable}
\end{table*}

\begin{table*}[h]
\centering
\begin{threeparttable}
\caption{Analysis of FLOP count in double-precision operations for forming the four-component Dirac-Coulomb-Gaunt Hamiltonian. $N$ number of atomic basis functions are used in this estimate. Restricted-kinetic-balanced (RKB) basis is compared with the unrestricted-kinetic-balanced (UKB) scalar basis. FLOP counts are separated into Coulomb $\mathbf{J}$ and exchange $\mathbf{K}$ contractions. Dense matrices with no integral or matrix symmetry are considered.}
\label{tab:FLOP}
\footnotesize

\begin{tabular*}{6.5in}{@{\extracolsep{\fill} } lcccc}
\hline

\hline
\multirow{2}{*}{Four-Component Scheme} & \multicolumn{3}{c}{Number of Integral-Density Contractions $(\mathbf{J},\mathbf{K})$} &\multirow{2}{*}{Total FLOP$^{[b]}$}\\
\cline{2-4}
 & 
 Dirac-Coulomb$^{[a]}$  & $(SS|SS)$ & Gaunt  &  \\ 
\hline

Pauli Matrix RKB Basis &  $(9, 20)$ & $(12,33)$ &$(28, 57)$ & $636N^4$\\
Two-Spinor RKB Basis &  $(48,48)$ & $(16, 16)$ &$(64,64)$ &$2048N^4$ \\
UKB Scalar Basis$^{[c]}$ &  $(54,54)$ & $(156, 156)$&$(25,100)$ &$2180N^4$  \\
\hline
\hline
\end{tabular*}

\begin{tablenotes}
\item[a] Dirac-Coulomb without the $(SS|SS)$ contribution.
\item[b] Floating point operation (FLOP) is defined as an arithmetic procedure ($+,-,\times,\div$) of two double-precision numbers. Because integrals for Pauli matrix RKB basis and UKB scalar basis are stored as real-valued matrices and density matrices are complex-valued, each contraction step requires 4$N^4$ FLOPs. For two-spinor RKB basis, because integrals are complex-valued, each contraction step requires 8$N^4$ FLOPs.
\item[c] For scalar-basis based approach, we assume there are $\frac{N}{2}$ number of $L>0$ atomic basis function in this set, resulting in a total of 2.5$N$ number of smaller component basis functions. For the estimation of operation count, see Appendix C.
\end{tablenotes}
\end{threeparttable}
\end{table*}

\Cref{tab:RAM} and \Cref{tab:FLOP} lists the estimated computational costs in terms of memory requirement and FLOP count for forming the Dirac-Fock matrix from assembled integrals, comparing several different four-component formalisms. In this estimate, we used a Kramers-unrestricted condition where all four Pauli components are non-zero. The cost analysis for UKB scalar basis is based on the assumption that 50\% of the basis functions have a non-zero angular momentum.

For all methods considered in \Cref{tab:RAM} and \Cref{tab:FLOP}, the computational cost of real-valued ERI and ERI derivatives are considered equivalent.
We also assume the computational cost of assembling the ERI and ERI derivatives into the corresponding two-spinor form is the same as assembling them into unique one-component integrals needed for the Pauli matrix formalism.
With these valid assumptions, the difference in computational cost for all four-component implementations comes from the different ways of computing the integral-density contractions in order to form the Dirac-Fock matrix.

For the Pauli matrix formalism presented in this work, and the method based on the UKB scalar basis, instead of assembling ERI and ERI derivatives into the two-spinor form, all integrals remain in the one-component form. As a result, for storing integrals in memory, the Pauli matrix  and  scalar basis formalisms have a smaller memory storage requirement as shown in \Cref{tab:RAM}. 

As discussed previously, the maximal component- and spin-separation in the Pauli spinor representation allows for combining all integral-density contractions of a same integral type. As a result, the four-component Dirac equation in the Pauli matrix representation is likely to be the method with minimal FLOP count. \Cref{tab:FLOP} shows the FLOP count for each type of ${\bf{J}}$ and ${\bf{K}}$ contractions with the density. For the Dirac-Coulomb terms we see that the Pauli matrix formalism has the minimum FLOP count for both ${\bf{J}}$ and ${\bf{K}}$ matrices compared to the two-spinor basis and the UKB scalar basis by a factor of $\frac{1}{2}$ or smaller. For the $(SS|SS)$ contribution, the ${\bf{J}}$ FLOP count of the Pauli matrix formalism is the minimum of the three, but the ${\bf{K}}$ FLOP count is more than the two-spinor formalism. The UKB scalar basis ends up being more than 5-10 times as expensive overall than the Pauli matrix, or two-spinor formalism. The UKB scalar basis method is particularly disadvantageous for the $(SS|SS)$ term simply because the large $N_S^4$ operation.
Finally, for the Gaunt term we see that the UKB scalar basis has the lowest number of FLOPs for the ${\bf{J}}$ contraction, but the Pauli matrix formalism has the lowest number for the ${\bf{K}}$ contraction. 

While there isn't one single four-component scheme that has the lowest number of FLOPs for every kind of contraction term, the Pauli matrix formalism is optimal in 4 out of 6 types of integral-density contractions considered here. We see that after including all of the contracted terms for the Dirac-Coulomb-Gaunt Hartree--Fock procedure, the Pauli matrix formalism has the best scaling FLOP prefactor by a factor of $\frac{1}{3}$ or smaller, where the two-spinor and UKB scalar form are comparable.

\subsection{Integral and Contraction Timings}
For the following timing benchmarks, we use a lanthanum dimer with a bond distance of 3.0 \AA \, and the  uncontracted ANO-RCC basis set.\cite{Widmark04_2851,Widmark05_6575,Borin08_11431} Computations were performed on an  Intel\textsuperscript{\textregistered} Xeon\textsuperscript{\textregistered} W 2.5 GHz CPU. Additionally, Chronus Quantum was compiled using the Clang Compiler version 12.0.0. Additionally, the integrals are computed using the LIBCINT library\cite{Sun15_1664}, with a development version of the Chronus Quantum electronic structure software package.\cite{Li20_e1436}  Note that the timings serve the purpose to show the relative costs of computationally dominant steps in Dirac Hartree--Fock build. They should not be used as calibrations of the performance of the program as the code optimization is beyond the scope of this article.

\begin{figure}[ht]
	\begin{center}
		\includegraphics[width=3.75in]{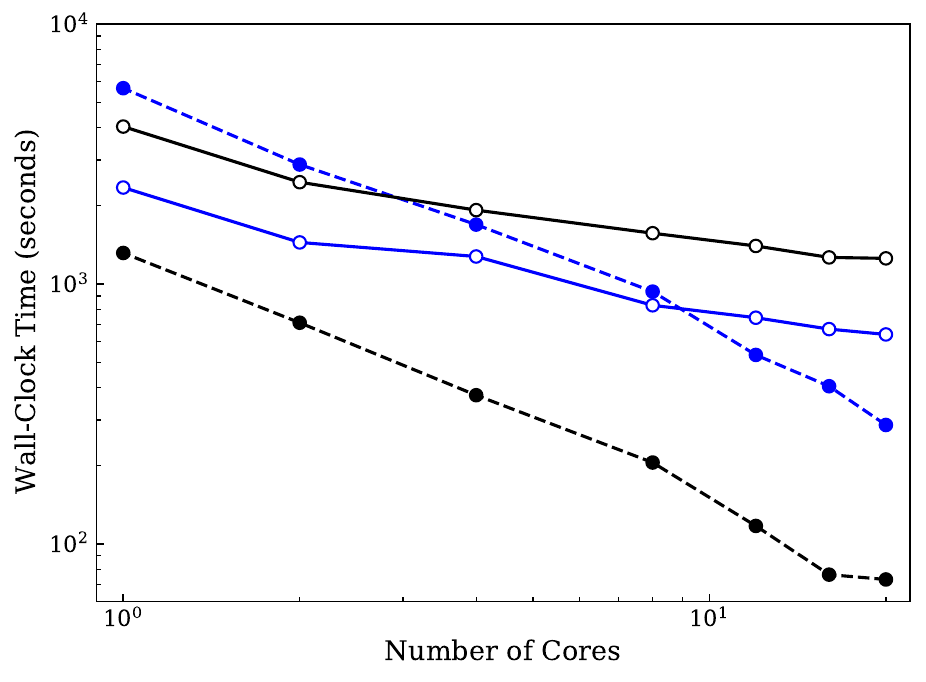}\\
		\caption{Wall-clock timings for the computation of integrals (solid lines with empty circles) and the duration of the AO-direct density-integral contraction (dashed lines with filled circles) for the lanthanum dimer using the uncontracted ANO-RCC basis. The time-reversal symmetry is turned off. The Schwarz integral screening threshold of $10^{-12}$ is used. The timings for the Dirac-Coulomb and the Gaunt two-body terms, shown in black and blue, respectively, were collected from a single DHF build. Logarithmic wall-clock times are plotted against logarithmic number of CPU cores. The lines are included in this plot are linear interpolations to guide the eye.}
		\label{fig:timings}
	\end{center}
\end{figure}

While \Cref{tab:FLOP} analyzes the computational cost at the theoretical limit in the absence of all symmetries, benchmark calculations presented here take advantage of integral symmetries and matrix hermiticity. Note that all calculations assume the Kramers' unrestricted case unless otherwise noted.
Although both the Dirac-Coulomb (without $(SS|SS)$) and Gaunt integrals are two-center second derivatives, the Gaunt integrals have a 2-fold symmetry whereas the integral symmetry is 4-fold for Dirac-Coulomb integrals (without $(SS|SS)$). The four-center derivatives of the $(SS|SS)$ term has an 8-fold symmetry.
\Cref{fig:timings} shows the wall-clock time for the lanthanum dimer system per number of cores and for each type of two-body integral and density contractions respectively, based on the Pauli spinor representation formalism presented here.  For DC-HF, the $(SS|SS)$ term is also included.

The ratio of the contraction times on a single CPU core for the Dirac-Coulomb and the Gaunt term is $\sim1:4.3$. This ratio is close to the theoretical limit of DCG in Pauli representation when integral and matrix symmetries are considered. The lower computational cost for DC contraction is mainly due to the higher symmetry of the underlying integrals.
The ratio of the computational cost for calculating the integrals needed for the Dirac-Coulomb and the Gaunt terms is $\sim1.7:1$. In addition to two-center second derivatives, the Dirac-Coulomb term also includes the four-center fourth-derivatives
of integrals arising from the $(SS|SS)$ contribution.
As a result, the Dirac-Coulomb integrals are more computationally expensive than those for the Gaunt term. 
The overall ratio of the computational cost, including both integral evaluation and contraction, for the Dirac-Coulomb and the Gaunt terms is $\sim1:1.5$ on a single CPU core. 

\Cref{fig:timings} also shows that on a single CPU core the cost of computing the Dirac-Coulomb term is dominated by the integral evaluations, whereas it is the contraction step that is the more expensive step for the Gaunt term. 
The timing ratios between the integral evaluation and contraction are $3:1$ and $1:2.4$ for the Dirac-Coulomb and the Gaunt terms, respectively, on a single CPU core.
The times for Dirac-Coulomb and Gaunt contractions decrease roughly $1/n$ for every increase in $n$ CPU cores used. The computational cost of calculating two-electron repulsion integrals and integral derivatives also decreases when increasing the number of CPU cores computing in parallel. The contraction step in AO-direct benefits more from the parallelization, as it is FLOP-bound. At around 8 cores, the cost of the Gaunt contraction decreases below the the cost of integral derivative evaluations.

\begin{figure}[htbp]
     \centering
            \includegraphics[width=.45\textwidth]{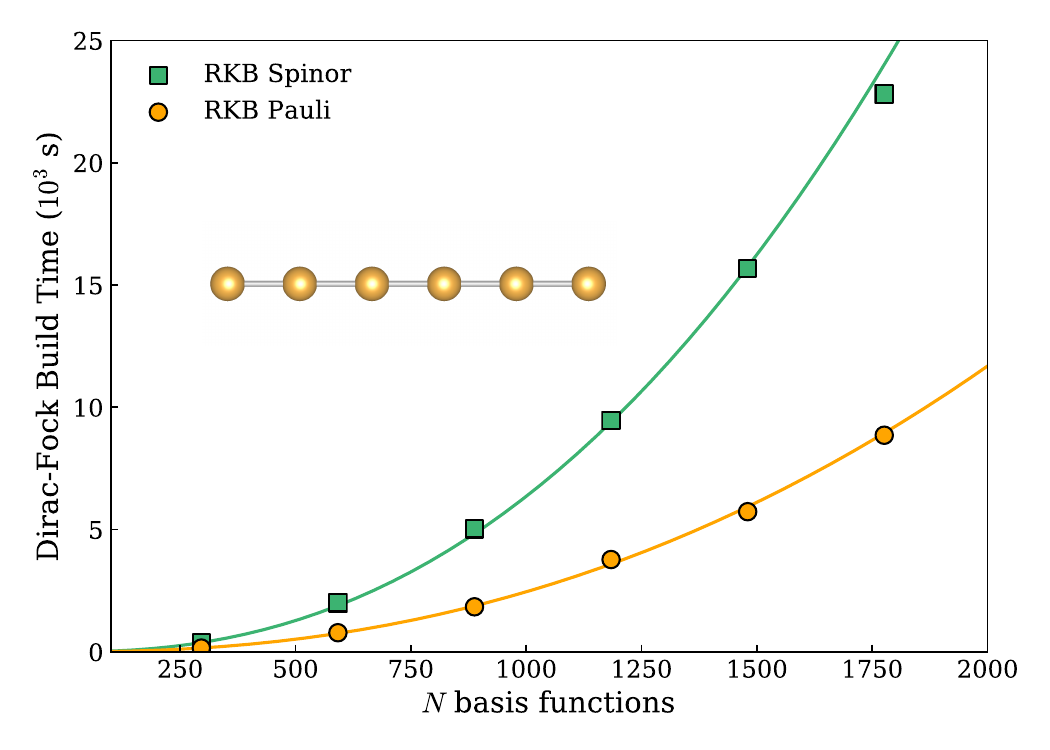}
            \includegraphics[width=.45\textwidth]{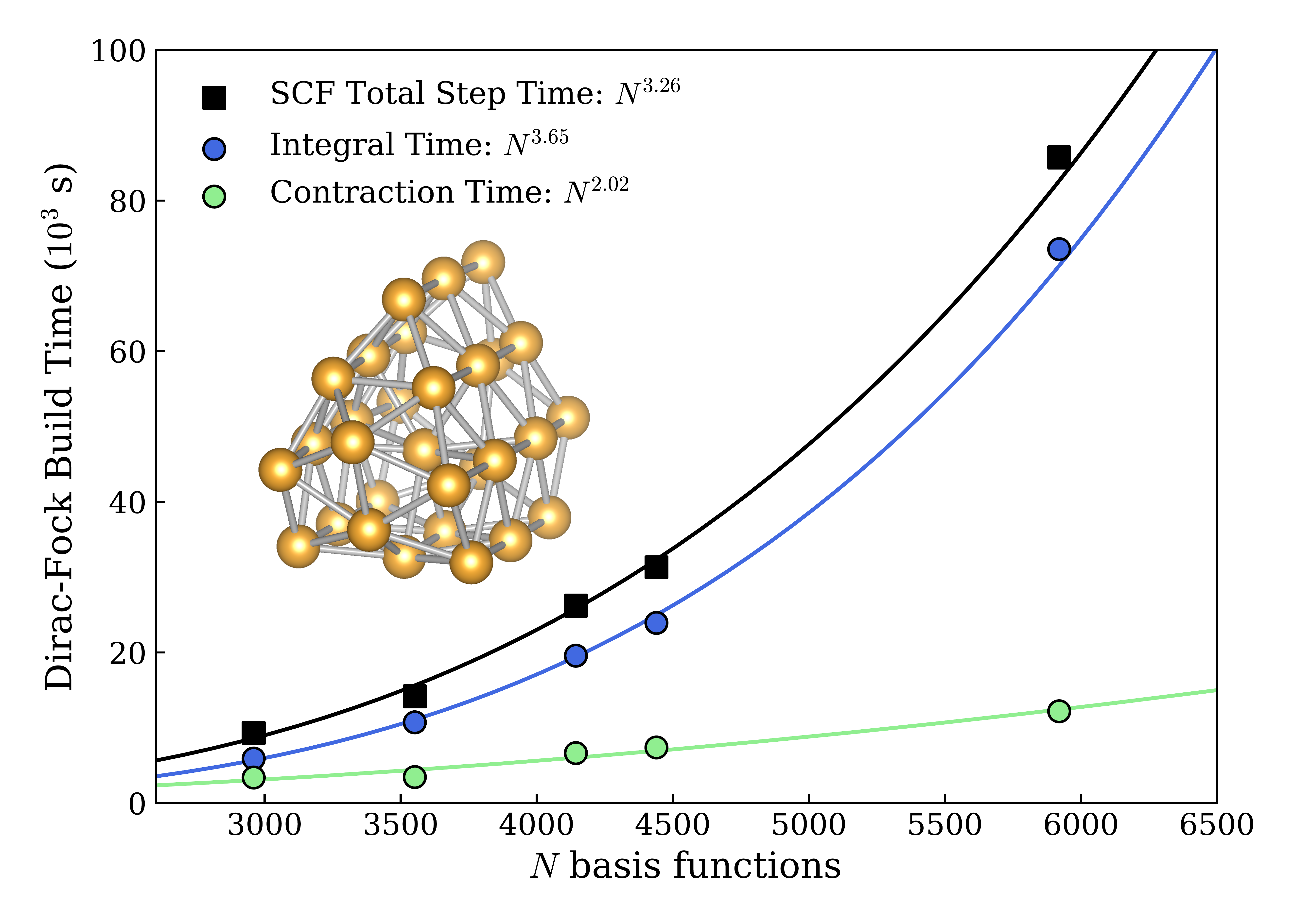}
		    \vspace{-10pt}
		    \caption{({\bf Left}) Single-core timing comparison for building the Dirac-Hartree-Fock matrix with the Dirac-Coulomb operator without the $(SS|SS)$ term using the RKB spinor and the RKB Pauli representation, respectively. A series of Au linear chains (Au$_2$ to Au$_{12}$) were used for this test with the uncontracted Jorge-DZ-DKH basis using an AO direct algorithm. A Schwarz screening threshold of $10^{-12}$ is used in each calculation. ({\bf Right}) 24-core parallel performance of the RKB Pauli Dirac-Hartree-Fock algorithm with the Dirac-Coulomb operator for building the Dirac-Hartree-Fock matrix. A series of Au clusters (Au$_{20}$ to Au$_{40}$) were used for this result. The RKB spinor results were computed using PySCF\cite{Chan18_e1340} and the RKB Pauli results were computed in ChronusQ.\cite{Li20_e1436}}\label{fig:Aucluster}
\end{figure}

Next, \cref{fig:Aucluster} shows the timing comparison for building the Dirac-Hartree-Fock matrix with the Dirac-Coulomb operator using the RKB spinor and RKB Pauli representations respectively. Computations were performed on an  Intel\textsuperscript{\textregistered} Xeon\textsuperscript{\textregistered} Platinum 8160 CPU. Additionally, Chronus Quantum was compiled using the Intel\textsuperscript{\textregistered} C++ Compiler version 19.0.1.144.  RKB spinor calculations were performed using PySCF.\cite{Chan18_e1340} In order to compare these implementations on the same footing, the calculations in our RKB spinor method for these results were computed utilizing time-reversal symmetry, and the contractions are done using an atomic-orbital direct scheme. The observed ratio of wall-clock times  between the two representations for linear Au chains agrees with the time complexity analysis in \cref{tab:FLOP}, where the RKB spinor representation is roughly 3 times slower than the RKB Pauli method. 

Additionally, we present timings for the Dirac-Fock build in our RKB Pauli representation in \cref{tab:FLOP}, split into both the density contraction and integral build times for a 24-core parallel calculation. To showcase this comparison we used bulky Au cluster calculations from Au$_{20}$ to Au$_{40}$. Since the RKB Pauli representation greatly reduces the density-integral contraction time, it can be seen that the computational cost of the integral build step dominates the overall computational cost. For a Dirac-Coulomb calculation with $\sim$6000 basis functions, the RKB Pauli based density contraction only takes a little more than 2 hours on a 24-core workstation (2.5 GHz Intel Xeon W).

Including relativistic two-electron contributions is by no means computationally cheap, especially including the Gaunt term which dominates the cost by being an order of magnitude greater than both the Dirac-Coulomb and Bare-Coulomb integrals. However, the method presented here is possibly the optimal algorithm for DHF. Integral and contraction acceleration techniques, such as the fast-multipole-method and density-fitting, can be applied to reduce the computational cost further based on the RKB Pauli formalism.

\subsection{Atomic Periodic Trend} \label{atomic}

In the data presented in this section, the integrals and the derivatives of integrals with respect to nuclear coordinates are computed with the LIBINT and LIBCINT libraries.\cite{Libint2,Sun15_1664} Additionally, the four-component Dirac Hartree--Fock formalism is implemented in a development version of the Chronus Quantum software package.\cite{Li20_e1436} The speed of light constant used in the following calculations is $137.03599967994$ a.u.

The main purpose of this paper is to introduce the methodological foundation and formalism of the four-component Dirac Hartree--Fock equation in the Pauli spinor representation. The four-component framework is validated by comparing the computed energies using the analytical results of the Dirac equation for hydrogenic atoms, shown in Appendix D, and basis set limit tests, shown in Appendix E. When a large uncontracted basis set is used, the calculations show a percent error on the order of $10^{-3} \%$. For benchmarking, we focus on analyzing and comparing the ground state DHF energies with different levels of relativistic two-electron interactions. In order to analyze the importance of relativistic two-electron corrections in Dirac Hartree--Fock, we introduce a Bare-Coulomb approximation in matrix form, \emph{i.e.} only the first term in \cref{eq:vcll} is included.

Several atomic series are selected from the periodic table, including the halogens, the coinage metals, and the lanthanides. In this way we can compare each relativistic two-body interaction contribution as a function of element charge, $Z$, although this separation of terms is not unique to the RKB Pauli formalism. In this section we report the ground state energies of these atoms using the uncontracted ANO-RCC basis set\cite{Widmark04_2851,Widmark05_6575,Borin08_11431}. The SCF energies for each atom with different levels of relativistic two-electron interaction are shown in \cref{table:anorccatoms} for each selected atom.

\begin{table}[htbp]
{
{\footnotesize
\caption{Ground state energies (Hartrees) for selected atoms with the energetic contributions of Dirac-Coulomb  and Gaunt terms, computed as $\Delta E^{DC} = E^{DC} - E^{BC}$ and $\Delta E^{G} = E^{DCG} - E^{DC}$ where $E^{DCG}$, $E^{DC}$, and $E^{BC}$ are the SCF energies of Dirac-Coulomb-Gaunt, Dirac-Coulomb, and Bare-Coulomb calculations respectively. The uncontracted ANO-RCC basis with finite width nuclei was used for this dataset. } 
\centering 
\begin{tabular}{c r r r r r} 
\hline
Atom & Bare-Coulomb & Dirac-Coulomb & $\Delta E^{DC}$ & Dirac-Coulomb-Gaunt & $\Delta E^{G}$\\ [0.5ex] 
\hline 
\multicolumn{2}{l}{Halogen Series}\\
F & $-99.540917$ & $-99.504738$ & $0.036179$ & $-99.492763$ & $0.011975$  \\ 
Cl & $-461.406677$ & $-460.947164$ & $0.459512$ & $-460.830018$ & $0.117146$ \\ 
Br & $-2612.852939$ & $-2605.033836$ & $7.819103$ & $-2603.603969$ & $1.429867$ \\ 
I & $-7155.395184$ & $-7115.798492$ & $39.596692$ & $-7109.770036$ & $6.028456$ \\ 
 [1ex] 
\hline
\multicolumn{2}{l}{Lanthanide Series}\\
La &  $-8546.215923$ & $-8493.657945$& $52.557978$ & $-8485.891031$ & $7.766914$ \\
Ce &  $-8917.383681$ & $-8861.043868$& $56.339814$ & $-8852.800219$ & $8.243649$ \\
Pr &  $-9298.457621$ & $-9238.164865$& $60.292756$ & $-9229.419749$ & $8.745116$ \\
Nd &  $-9689.682825$ & $-9625.317052$& $64.365773$ & $-9616.050622$ & $9.266430$ \\
Pm &  $-10091.178921$ & $-10022.330954$& $68.847967$ & $-10012.520248$ & $9.810706$ \\
Sm &  $-10502.861365$ & $-10429.356216$& $73.505149$ & $-10418.975555$ & $10.380661$ \\
Eu &  $-10925.261820$ & $-10846.946858$& $78.314963$ & $-10835.981295$ & $10.965563$ \\
Gd &  $-11358.014014$ & $-11274.744433$& $83.269580$ & $-11263.160904$ & $11.583529$ \\
Tb &  $-11801.625306$ & $-11712.825124$& $88.800182$ & $-11700.608150$ & $12.216974$ \\
Dy &  $-12256.214858$ & $-12161.776442$& $94.438417$ & $-12148.896505$ & $12.879937$ \\
Ho &  $-12721.922596$ & $-12621.573979$& $100.348617$ & $-12608.005042$ & $13.568938$ \\
Er &  $-13198.887030$ & $-13092.441227$& $106.445803$ & $-13078.151685$ & $14.289542$ \\
Tm &  $-13687.098882$ & $-13574.398156$& $112.700726$ & $-13559.365996$ & $15.032160$ \\
Yb &  $-14187.311321$ & $-14067.663235$& $119.648086$ & $-14051.865403$ & $15.797831$ \\
Lu &  $-14699.426395$ & $-14572.526419$& $126.899976$ & $-14555.924058$ & $16.602362$ \\
 [1ex] 
\hline
\multicolumn{2}{l}{Coinage Metal Series}\\
Cu & $-1657.190117$ & $-1653.453398$ & $3.736719$ & $-1652.708252$ & $0.745146$   \\
Ag & $-5339.393098$ & $-5314.625699$ & $24.767399$ & $-5310.654463$ & $3.971236$   \\
Au & $-19231.541954$ & $-19035.579049$ & $195.962905$ & $-19011.375550$ & $24.203499$   \\
[1ex] 
\hline 
\end{tabular}
\label{table:anorccatoms} 
}
}
\end{table}

\begin{figure}[htbp]
	\begin{center}
		\includegraphics[width=3.75in]{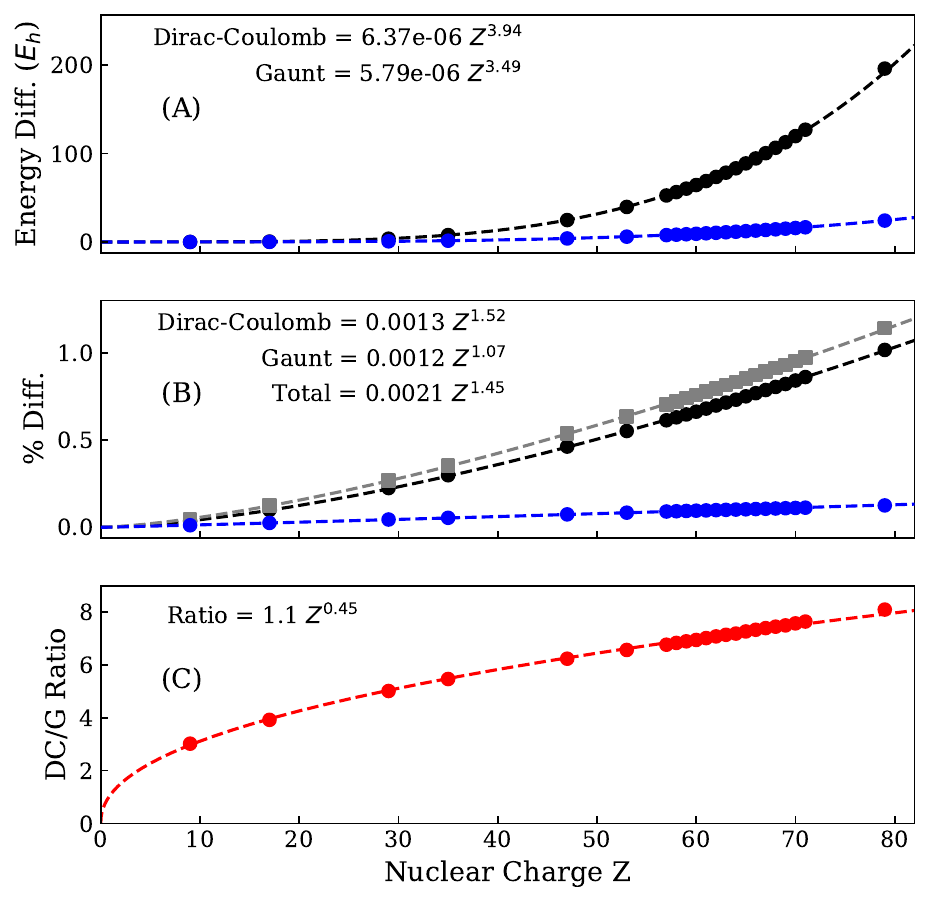}\\
		\caption{(A) Energetic contributions of Dirac-Coulomb (black) and Gaunt (blue) terms, computed as $\Delta E^{DC} = E^{DC} - E^{BC}$ and $\Delta E^{G} = E^{DCG} - E^{DC}$ where $E^{DCG}$, $E^{DC}$, and $E^{BC}$ are the SCF energies of Dirac-Coulomb-Gaunt, Dirac-Coulomb, and Bare-Coulomb calculations. (B) Percentage of energy contribution of each two-electron term compared to the Bare-Coulomb ground state DHF energy, presented as $\frac{\Delta E^{DC}}{E^{BC}}\times100\%$ and $\frac{\Delta E^{G}}{E^{BC}}\times100\%$. The gray squares correspond to the total relativistic contribution which is $\frac{\Delta E^{DC} + \Delta E^{G}}{E^{BC}}\times100\%$ (C) The ratio of the energy contribution of the Dirac-Coulomb term vs. the Gaunt term $\frac{\Delta E^{DC}}{E^{G}}$ as a function of increasing nuclear charge for the neutral atom series. The power law fit is displayed as the dashed line, corresponding to the equation in the top left corner of each panel.}
		\label{fig:energy_diff}
	\end{center}
\end{figure}

From the computed data, we see that with each successive correction to the bare non-relativistic Coulomb interaction, the ground state energy increases. This is  expected as they all arise from the $\frac{1}{r_{12}}$ two-electron repulsion interaction. The relativistic two-electron interaction increases much faster as $Z$ increases in the same element group down the periodic table than the trend across a same row (such as the lanthanides). By comparing Dirac-Coulomb calculations using the uncontracted ANO-RCC basis with the previously published numerically exact values,\cite{Clementi90_1829,Dyall97_207} the observed errors are on the order of $10^{-7} \%$ to $10^{-5} \%$, showing excellent agreement.

The periodic trends of the relativistic two-electron interactions are plotted in \cref{fig:energy_diff}. For each data set, a least squares power law fit was applied and displayed as the dashed lines for each subplot in \cref{fig:energy_diff}.
\Cref{fig:energy_diff}A shows that the Dirac-Coulomb and Gaunt contributions to the two-body interaction increase polynomially as $Z$ increases. The Dirac-Coulomb term exhibits a near quartic growth with respect to $Z$, while the Gaunt term has a $Z^{3.49}$ dependence. It is more meaningful to analyze the energetic contributions in terms of the percentage of the total energy. In \cref{fig:energy_diff}B, the percent contribution is evaluated with respect to the Bare-Coulomb DHF energy. This plot compares the growth behaviors of relativistic two-electron interactions relative to that exhibited by the Bare-Coulomb ground state energy.
It is quite surprising to see that the Gaunt contribution in terms of the percent of the total energy is almost linear with respect to $Z$, suggesting that the relativistic Gaunt interaction behaves almost like the Bare-Coulomb term. \Cref{fig:energy_diff}B also shows that the Dirac-Coulomb contribution exhibits a much faster growth behavior than both the Gaunt and the Bare-Coulomb interaction. As a result, \cref{fig:energy_diff}C shows that the the Dirac-Coulomb interaction is a factor of 3$-$5 times larger than the Gaunt contribution for light elements ($Z<40$), and increases towards a ratio of $\sim$9 near the end of the periodic table. As $Z$ increases, this ratio increases as $\sqrt{Z}$ approximately.

\subsection{Coinage Metal Dimers}
In this section, the equilibrium bond lengths and dissociation energies for the coinage metal dimer series including Cu$_2$, Ag$_2$, Au$_2$, are computed. The uncontracted ANO-RCC basis set\cite{Widmark05_6575} is used for all calculations throughout this section. For these results we would like to emphasize that agreement with experimental data should not be expected due to the lack of electronic correlation.\cite{Schwerdtfeger99_9457} Analyses presented here aim to provide a qualitative understanding of how the Dirac-Coulomb and Gaunt two-electron relativistic effects change molecular properties compared to the non-relativistic results. 

Potential energy scans are carried out using the non-relativistic Hartree-Fock (NR-HF), Dirac-Coulomb, and Dirac-Coulomb-Gaunt Hamiltonians to find the equilibrium bond length for each dimer and level of theory, reported in \cref{table:dimer_bonds}. Additionally, the bond dissociation energy ($D_e$) is computed as $D_e = 2E_{\text{atom}} - E_{\text{dimer}}$, where $E_{\text{atom}}$ is the atomic energy and $E_{\text{dimer}}$ is the energy of the dimer at the equilibrium bond length. The theoretical and experimental results for the dissociation energies are presented in \Cref{table:dimer_de}.

\begin{table}
\caption{The equilibrium bond lengths for each dimer system are reported in \AA. The experimental bond lengths are reported under ``Exp." The uncontracted ANO-RCC  basis is used for all calculations.} 
\centering 
\begin{tabular}{c r r r r} 
\hline
Dimer & NR-HF & Dirac-Coulomb & Dirac-Coulomb-Gaunt & Exp. \\ [0.5ex] 
\hline 
Cu$_2$ &  2.409 & 2.402 & 2.403 & 2.22 [Ref. \citenum{Morse86_1049}] \\
Ag$_2$ &  2.770 & 2.701 & 2.703 & 2.53 [Ref. \citenum{Langridge-Smith91_415}]\\
Au$_2$ &  2.830 & 2.586 & 2.590 & 2.47 [Ref. \citenum{Barrow67_39}]
\\[1ex] 
\hline 
\end{tabular}
\label{table:dimer_bonds} 
\end{table}

\begin{table}
\caption{Equilibrium bond dissociation energies ($D_e$) for each dimer system are reported in eV. The experimental dissociation energies are reported under ``Exp." The uncontracted ANO-RCC  basis is used for all calculations.} 
\centering 
\begin{tabular}{c r r r r} 
\hline
Dimer & NR-HF & Dirac-Coulomb & Dirac-Coulomb-Gaunt & Exp. \\ [0.5ex] 
\hline 
Cu$_2$ &  0.600 & 0.549 & 0.547 & 2.01$\pm$0.08 [Ref. \citenum{Morse86_1049}] \\
Ag$_2$ &  0.486 & 0.480 & 0.477 & 1.65$\pm$0.03 [Ref. \citenum{Gingerich80_739}] \\
Au$_2$ &  0.659 & 0.865 & 0.855 & 2.29$\pm$0.02 [Ref. \citenum{Seyse74_5114}] 
\\[1ex] 
\hline 
\end{tabular}
\label{table:dimer_de} 
\end{table}

\Cref{table:dimer_bonds} shows that the Dirac-Coulomb computed bond length contracts for all dimers studied here relative to the non-relativistic result. This is mainly due to the core-electron contraction arising from the scalar relativistic effects.\cite{Schwarz02_804} The main takeaway from the equilibrium bond length calculations in \cref{table:dimer_bonds} is that the Gaunt correction to the Dirac-Coulomb two-electron interaction introduces an additional electron-electron repulsion that elongates the bond length by 0.001--0.004\AA. In \cref{table:dimer_de} the Gaunt correction shows a systematic trend that lowers the bond dissociation energy on the order of $\sim$10 meV. This observation suggests that the Gaunt term gives rise to a small yet increasingly important repulsive correction to the bond dissociation energy as the atomic number increases. 

\section{Conclusion and Perspective}

In this work, we introduced a computationally efficient four-component Dirac Hartree--Fock (DHF)
formalism within the maximally component- and spin-separated Pauli spinor representation in the restricted-kinetic-balance condition (denoted ``RKB Pauli'').
A minimal FLOP count algorithm is developed for the density contraction with the relativistic electron repulsion integrals. All of the required ingredients for this implementation can utilize the non-relativistic atomic orbital electron repulsion integrals and their gradients stored in the one-component form without going through the two-spinor basis construction. Additionally, a detailed cost analysis was presented and compared to other four-component algorithms. 
 Numerical tests shows that the RKB Pauli representation is physically equivalent to the RKB spinor representation, but computationally much cheaper.

Additionally, the contribution of relativistic two-electron interactions across the periodic table were analyzed. As expected, both Dirac-Coulomb and Gaunt terms increase the total energy, with the Dirac-Coulomb contribution exhibiting a faster growth as $Z$ increases. The importance of the Gaunt interaction to the total atomic energy increases linearly with respect to the non-relativistic Coulomb contribution while the Dirac-Coulomb term becomes more important relative to the Gaunt term as $Z$ increases.

The RKB Pauli formalism of the four-component DHF Hamiltonian presented in this work provides a computationally efficient way to develop relativistic electronic structure theories. Although the final mathematical working expressions are more complicated than those in two-spinor or scalar basis, we are rewarded with an optimal algorithm with a minimal memory requirement and total FLOP count. This work lays the computational foundation for practical applications of four-component DHF, and future developments of correlated wave function methods in the four-component domain.

When it comes to computing bond lengths and dissociation energies for metal dimers, our results show that the Gaunt correction introduces an additional electron-electron repulsion that slightly elongates the bond length and increases the  bond dissociation energy. Additionally, as expected, electron correlation is vital for making quantitative comparisons with experimental values.

For this study, we primarily focused on comparing the computational performance of DHF methods utilizing fully uncontracted basis sets.  While introducing  contracted basis sets can reduce the resulting DHF matrix size, using \emph{standard} contracted basis sets in 4-component DHF is not recommended.\cite{Faegri01_252,Nieuwpoort91_131}
Naively fixing the contraction coefficients for both the large and small components of the wave function can lead to variational collapse.\cite{Nieuwpoort91_131,Liu10_1679} One approach to mitigate this problem is to use the atomic balance basis, proposed and pioneered by Visscher and Liu,\cite{Nieuwpoort91_131,Liu10_1679} which allows the small and large component bases to have their own contraction coefficients derived from an atomic calculation with the uncontracted basis. However, the best way to extend the present method to this case is currently under investigation.

\begin{acknowledgement}
We would like to thank Prof. Trond Saue for helpful discussions on various kinetic balance conditions in Dirac Hartree-Fock theory. Additionally, X. L. acknowledges support from the U.S. Department of Energy, Office of Science, Basic Energy Sciences, in the Heavy-Element Chemistry program (Grant No. DE-SC0021100) for the development of electronic structure methods within the four-component Dirac framework. The development of the open source software package is supported by the U.S. National Science Foundation (OAC-1663636). The work of E. F. V. on the evaluation of electron repulsion integrals and integral derivatives was supported by the U.S. National Science Foundation (Grant Nos. 1550456 and 1800348).
\end{acknowledgement}

\subsection*{Appendix A: Spin-Separation for Gaunt Interaction}

The Gaunt term integrals can be separate into spin-free, spin-own-orbit, and spin-other-orbit interactions in the Pauli spinor representation.
In this section, detailed derivations will be shown for the $(\mu^L\boldsymbol\sigma \nu^S|\cdot \kappa^S\boldsymbol\sigma \lambda^L)$ type of integral which contributes to the exchange term in $V^{G,LL}$ and Coulomb term in $V^{G,LS}$.

The $(\mu^L\boldsymbol\sigma \nu^S|\cdot \kappa^S\boldsymbol\sigma \lambda^L)$ integral can be written as (prefactor $\frac{1}{(2mc)^2}$ is ignored for brevity),
\begin{align}
	(\mu^L _{\tau_1} \boldsymbol\sigma \nu^S _{\tau_2}| \cdot \kappa ^S_{\tau_3} \boldsymbol\sigma \lambda ^L _{\tau_4}) 
	=&
	(\mu^L _{\tau_1} \boldsymbol\sigma(1) \boldsymbol\sigma(1)\cdot \mathbf{p}(1) \nu^L _{\tau_2} | \cdot \boldsymbol\sigma(2) \cdot \mathbf{p}(2) \kappa ^L_{\tau_3} \boldsymbol\sigma(2) \lambda ^L _{\tau_4})\nonumber \\
	=&-(\mu^L _{\tau_1}\boldsymbol\sigma(1) \boldsymbol\sigma(1)\cdot \boldsymbol\nabla(1) \nu ^L _{\tau_2}| \cdot \boldsymbol\sigma(2) \cdot \boldsymbol\nabla(2) \kappa ^L _{\tau_3} \boldsymbol\sigma(2) \lambda ^L _{\tau_4}) \nonumber
\end{align}
where we used the restricted-kinetic-balance condition $\mu^S = \frac{1}{2mc}\boldsymbol\sigma\cdot {\bf p} \mu^L$. 
$\boldsymbol\nabla_\kappa$ and $\boldsymbol\nabla_\nu$ are gradients with respect to basis centers of $\kappa^L$ and $\nu^L$, respectively. 

We now switch from Mulliken notation to Dirac notation to carry out the spin-separation
\begin{align}
	&-(\mu^L _{\tau_1}\boldsymbol\sigma(1) \boldsymbol\sigma(1)\cdot \boldsymbol\nabla(1) \nu ^L _{\tau_2}| \cdot \boldsymbol\sigma(2) \cdot \boldsymbol\nabla(2) \kappa ^L _{\tau_3} \boldsymbol\sigma(2) \lambda ^L _{\tau_4})\notag\\
	=&-\bra{\mu^L _{\tau_1}\boldsymbol\sigma(2)\cdot \boldsymbol\nabla_\kappa \kappa^L_{\tau_3}}  \frac{\boldsymbol\sigma(1)\cdot \boldsymbol\sigma(2)}{r_{12}} \boldsymbol\sigma(1)\cdot \boldsymbol\nabla_\nu \ket{\nu^L_{\tau_2} \lambda^L _{\tau_4}}\notag\\
	=&\bra{\mu^L _{\tau_1} \kappa^L_{\tau_3}} \boldsymbol\sigma(2)\cdot \boldsymbol\nabla_\kappa \frac{\boldsymbol\sigma(1)\cdot \boldsymbol\sigma(2)}{r_{12}} \boldsymbol\sigma(1)\cdot \boldsymbol\nabla_\nu \ket{\nu^L_{\tau_2} \lambda^L _{\tau_4}}\label{eq:Diracnotation}
\end{align}
where we used the relationship $\boldsymbol\nabla(2) = -\boldsymbol\nabla_\kappa$ between basis function derivative with respect to the electronic coordinate $\boldsymbol\nabla(2)$ and with respect to basis function center $\boldsymbol\nabla_\kappa$.

By using the Dirac identity, $(\boldsymbol\sigma(1)\cdot\boldsymbol\sigma(2))(\boldsymbol\sigma(1)\cdot\boldsymbol\nabla_\nu)=\mathbf{I}(1)~\boldsymbol\sigma(2)\cdot\boldsymbol\nabla_\nu+i\boldsymbol\sigma(1)\cdot\boldsymbol\sigma(2)\times\boldsymbol\nabla_\nu$, \cref{eq:Diracnotation} can be written as,
\begin{align}
	&\bra{\mu^L _{\tau_1}\kappa^L_{\tau_3}} \boldsymbol\sigma(2)\cdot \boldsymbol\nabla_\kappa \frac{\boldsymbol\sigma(1)\cdot \boldsymbol\sigma(2)}{r_{12}} \boldsymbol\sigma(1)\cdot \boldsymbol\nabla_\nu \ket{\nu^L _{\tau_2} \lambda^L_{\tau_4}}\notag\\
	=& \bra{\mu^L _{\tau_1}\kappa^L_{\tau_3}} \boldsymbol\sigma(2)\cdot \boldsymbol\nabla_\kappa
	\frac{1}{r_{12}}  
	[{\bf I}(1) \boldsymbol{\sigma}(2) \cdot \boldsymbol{\nabla}_\nu +i \boldsymbol{\sigma}(1) \cdot \boldsymbol{\sigma}(2)\times \boldsymbol{\nabla}_\nu]
	\ket{\nu^L _{\tau_2} \lambda^L_{\tau_4}}.\label{eq:separation1}
\end{align}
Additional applications of the Dirac identity are carried out to further spin-separate the two terms in \cref{eq:separation1}:
\begin{align}
    &\bra{\mu^L _{\tau_1}\kappa^L_{\tau_3}} \boldsymbol\sigma(2)\cdot \boldsymbol\nabla_\kappa
	\frac{1}{r_{12}}  
	{\bf I}(1) \boldsymbol{\sigma}(2) \cdot \boldsymbol{\nabla}_\nu
	\ket{\nu^L _{\tau_2} \lambda^L_{\tau_4}}\notag\\
	=&\bra{\mu^L _{\tau_1}\kappa^L_{\tau_3}}
	{\bf I}(1)\frac{1}{r_{12}}  
	[{\bf I}(2) \boldsymbol{\nabla}_\kappa \cdot \boldsymbol{\nabla}_\nu 
	+i \boldsymbol{\sigma}(2)\cdot \boldsymbol{\nabla}_\kappa \times \boldsymbol{\nabla}_\nu]
	\ket{\nu^L _{\tau_2} \lambda^L_{\tau_4}} \label{eq:separation2}
\end{align}
and
\begin{align}
    &\bra{\mu^L _{\tau_1}\kappa^L_{\tau_3}} \boldsymbol\sigma(2)\cdot \boldsymbol\nabla_\kappa
	\frac{1}{r_{12}}  
	i \boldsymbol{\sigma}(1) \cdot \boldsymbol{\sigma}(2)\times \boldsymbol{\nabla}_\nu
	\ket{\nu^L _{\tau_2} \lambda^L_{\tau_4}} \nonumber \\
	=&\bra{\mu^L _{\tau_1}\kappa^L_{\tau_3}}
	i\frac{1}{r_{12}}[
	{\bf I}(2) \boldsymbol{\nabla}_\kappa \cdot (\boldsymbol{\nabla}_\nu \times \boldsymbol{\sigma}(1)) + i \boldsymbol{\sigma}(2) \cdot \boldsymbol{\nabla}_\kappa \times 
	(\boldsymbol{\nabla}_\nu \times \boldsymbol{\sigma}(1)) ]
	\ket{\nu^L _{\tau_2} \lambda^L_{\tau_4}} \notag \\ 
	=&\bra{\mu^L _{\tau_1}\kappa^L_{\tau_3}}
	\frac{1}{r_{12}}[i
	{\bf I}(2)\boldsymbol{\sigma}(1) \cdot (\boldsymbol{\nabla}_\kappa \times \boldsymbol{\nabla}_\nu ) +  (\boldsymbol{\sigma}(2) \times \boldsymbol{\nabla}_\kappa) \cdot
	(\boldsymbol{\sigma}(1)  \times \boldsymbol{\nabla}_\nu ) ]
	\ket{\nu^L _{\tau_2} \lambda^L_{\tau_4}}
	\label{eq:separation3}
\end{align}

Combining \cref{eq:separation2} and \cref{eq:separation3} into \cref{eq:separation1}, and write the integrals in Mulliken notation, we arrive at the following expression:
\begin{align}
	(\mu^L _{\tau_1} \boldsymbol\sigma \nu^S _{\tau_2}| \cdot \kappa ^S_{\tau_3} \boldsymbol\sigma \lambda ^L _{\tau_4})=&~\xi_{\tau_1}^\dagger (1) \xi_{\tau_2} (1) \xi_{\tau_3}^\dagger (2) \xi_{\tau_4} (2) \big[\mathbf{I}(1)\mathbf{I}(2) \boldsymbol\nabla_\kappa\cdot\boldsymbol\nabla_\nu  \nonumber \\
	&+i (\mathbf{I}(1)\boldsymbol\sigma(2)+\mathbf{I}(2)\boldsymbol\sigma(1)) \cdot \boldsymbol\nabla_\kappa \times \boldsymbol\nabla_\nu \notag\\
	&+(\boldsymbol\sigma(2) \times \boldsymbol\nabla_\kappa) \cdot (\boldsymbol\sigma(1)\times \boldsymbol\nabla_\nu)\big](\mu\nu|\kappa\lambda)\label{eq:separation4}
\end{align}

In \cref{eq:separation4}, the integral type of $\mathbf{I}(1)\mathbf{I}(2) \boldsymbol\nabla_\kappa\cdot\boldsymbol\nabla_\nu$  gives rise to the spin-free term. The contribution from the integral type of $\mathbf{I}(1)\boldsymbol\sigma(2) \cdot \boldsymbol\nabla(2) \frac{1}{r_{12}}\times  \boldsymbol\nabla(1)$ is the spin-other-orbit term, while $\mathbf{I}(2)\boldsymbol\sigma(1) \cdot \boldsymbol\nabla(2) \frac{1}{r_{12}}\times \boldsymbol\nabla(1)$ is the spin-own-orbit term. 

The last term $(\boldsymbol\sigma(2) \times \boldsymbol\nabla_\kappa) \cdot (\boldsymbol\sigma(1)\times \boldsymbol\nabla_\nu)$ is the spin-spin interaction which can be further separated using the scalar quadruple product rule,
\begin{align}
    (\boldsymbol\sigma(2)\times\boldsymbol\nabla_\kappa)\cdot(\boldsymbol\sigma(1)\times\boldsymbol\nabla_\nu)&=(\boldsymbol\sigma(1)\cdot\boldsymbol\sigma(2))(\boldsymbol\nabla_\kappa\cdot\boldsymbol\nabla_\nu)-(\boldsymbol\sigma(1)\cdot\boldsymbol\nabla_\kappa)(\boldsymbol\sigma(2)\cdot\boldsymbol\nabla_\nu).\notag
\end{align}
The final, completely spin-separated expression for the $(\mu^L _{\tau_1} \boldsymbol\sigma \nu^S _{\tau_2}| \cdot \kappa ^S_{\tau_3} \boldsymbol\sigma \lambda ^L _{\tau_4})$ integral is
\begin{align}
	(\mu^L _{\tau_1} \boldsymbol\sigma \nu^S _{\tau_2}| \cdot \kappa ^S_{\tau_3} \boldsymbol\sigma \lambda ^L _{\tau_4})=&~\xi_{\tau_1}^\dagger (1) \xi_{\tau_1} (1) \xi_{\tau_3}^\dagger (2) \xi_{\tau_4} (2) \big[\mathbf{I}(1)\mathbf{I}(2) \boldsymbol\nabla_\kappa\cdot\boldsymbol\nabla_\nu  \nonumber \\
	&+i (\mathbf{I}(1)\boldsymbol\sigma(2)+\mathbf{I}(2)\boldsymbol\sigma(1)) \cdot \boldsymbol\nabla_\kappa \times \boldsymbol\nabla_\nu \notag\\ 
	&+(\boldsymbol\sigma(1)\cdot\boldsymbol\sigma(2)) (\boldsymbol\nabla_\kappa\cdot\boldsymbol\nabla_\nu) - (\boldsymbol\sigma(1)\cdot \boldsymbol\nabla_\kappa) (\boldsymbol\sigma(2)\cdot\boldsymbol\nabla_\nu) \big]
(\mu\nu|\kappa\lambda)
\label{eq:2spinorLSSLAPP}
\end{align}

\subsection*{Appendix B: Contributions from $(SS|SS)$}

The $SS$ block of the Dirac-Coulomb matrix has contributions from $(p^Sq^S|r^Ss^S)$ types of integral with a prefactor of $\frac{1}{(2mc)^4}$, denoted as $V^{C(4)}$. 
\begin{equation}
V^{C(4),SS}_{\mu\tau_1,\nu\tau_2}=\sum_{\lambda\tau_4\kappa\tau_3}D_{\lambda\tau_4\kappa\tau_3}^{SS}[(\mu^S_{\tau_1}\nu^S_{\tau_2}|\kappa^S_{\tau_3}\lambda^S_{\tau_4})-(\mu^S_{\tau_1}\lambda^S_{\tau_4}|\kappa^S_{\tau_3}\nu^S_{\tau_2})]
\end{equation}
These contributions are collectively known as the $(SS|SS)$ term, which is often ignored due to its large computational cost and small energetic contribution. The $(SS|SS)$ contribution in the Pauli spinor representation is:
\begin{align}
    V^{C(4), SS}_{\mu\nu , s} &=2\sum_{\lambda\kappa}  \bigg[ D_{\lambda\kappa, s}^{SS} (\boldsymbol{\nabla}_\mu \cdot \boldsymbol{\nabla}_\nu)(\boldsymbol{\nabla}_\kappa \cdot \boldsymbol{\nabla}_\lambda) + i \sum_{J=x,y,z} D_{\lambda\kappa,J}^{SS} (\boldsymbol{\nabla}_\mu \cdot \boldsymbol{\nabla}_\nu) (\boldsymbol{\nabla}_\kappa\times\boldsymbol{\nabla}_\lambda)_J \bigg] (\mu\nu|\kappa\lambda)\nonumber \\
    & - \sum_{\lambda\kappa} D_{\lambda\kappa, s}^{SS} \bigg[(\boldsymbol{\nabla}_\mu \cdot \boldsymbol{\nabla}_\lambda)( \boldsymbol{\nabla}_\kappa \cdot \boldsymbol{\nabla}_\nu)-\sum_{J=x,y,z} (\boldsymbol{\nabla}_\mu \times \boldsymbol{\nabla}_\lambda)_J( \boldsymbol{\nabla}_\kappa \times \boldsymbol{\nabla}_\nu)_J\bigg] (\mu\lambda|\kappa\nu) \nonumber\\
    &-i \sum_{\lambda\kappa}\sum_{J=x,y,z} D_{\lambda\kappa,J}^{SS}\bigg[(\boldsymbol{\nabla}_\mu \times \boldsymbol{\nabla}_\lambda)_J( \boldsymbol{\nabla}_\kappa \cdot \boldsymbol{\nabla}_\nu)+(\boldsymbol{\nabla}_\mu \cdot \boldsymbol{\nabla}_\lambda)( \boldsymbol{\nabla}_\kappa \times \boldsymbol{\nabla}_\nu)_J\bigg] (\mu\lambda|\kappa\nu) \nonumber\\
	&-i \sum_{\lambda\kappa} D_{\lambda\kappa, z}^{SS} \bigg[(\boldsymbol{\nabla}_\mu \times \boldsymbol{\nabla}_\lambda)_x( \boldsymbol{\nabla}_\kappa \times \boldsymbol{\nabla}_\nu)_y
	-  (\boldsymbol{\nabla}_\mu \times \boldsymbol{\nabla}_\lambda)_y( \boldsymbol{\nabla}_\kappa \times \boldsymbol{\nabla}_\nu)_x \bigg] (\mu\lambda|\kappa\nu)\nonumber \\
	& -i \sum_{\lambda\kappa}  D_{\lambda\kappa, y}^{SS} \bigg[(\boldsymbol{\nabla}_\mu \times \boldsymbol{\nabla}_\lambda)_z( \boldsymbol{\nabla}_\kappa \times \boldsymbol{\nabla}_\nu)_x 
	- (\boldsymbol{\nabla}_\mu \times \boldsymbol{\nabla}_\lambda)_x( \boldsymbol{\nabla}_\kappa \times \boldsymbol{\nabla}_\nu)_z \bigg] (\mu\lambda|\kappa\nu)\nonumber \\
	&
	-i \sum_{\lambda\kappa} D_{\lambda\kappa, x}^{SS}\bigg[ (\boldsymbol{\nabla}_\mu \times \boldsymbol{\nabla}_\lambda)_y( \boldsymbol{\nabla}_\kappa \times \boldsymbol{\nabla}_\nu)_z 
	-  (\boldsymbol{\nabla}_\mu \times \boldsymbol{\nabla}_\lambda)_z( \boldsymbol{\nabla}_\kappa \times \boldsymbol{\nabla}_\nu)_y  
	\bigg] (\mu\lambda|\kappa\nu) 
\end{align}

\begin{align}
    V^{C(4), SS}_{\mu\nu, z} &=
    2\sum_{\lambda\kappa}  \bigg[i
	D^{SS}_{\lambda\kappa, s} (\boldsymbol{\nabla}_\mu \times \boldsymbol{\nabla}_\nu)_z (\boldsymbol{\nabla}_\kappa \cdot \boldsymbol{\nabla}_\lambda)
	 -\sum_{J=x,y,z} D_{\lambda\kappa,J}^{SS} (\boldsymbol{\nabla}_\mu \times \boldsymbol{\nabla}_\nu)_z (\boldsymbol{\nabla}_\kappa\times\boldsymbol{\nabla}_\lambda)_J \bigg] (\mu\nu|\kappa\lambda)\nonumber \\
	 &- \sum_{\lambda\kappa} D_{\lambda\kappa, z}^{SS} \bigg[(\boldsymbol{\nabla}_\mu \cdot \boldsymbol{\nabla}_\lambda)( \boldsymbol{\nabla}_\kappa \cdot \boldsymbol{\nabla}_\nu)-(\boldsymbol{\nabla}_\mu \times \boldsymbol{\nabla}_\lambda)_z( \boldsymbol{\nabla}_\kappa \times \boldsymbol{\nabla}_\nu)_z\bigg] (\mu\lambda|\kappa\nu)\nonumber \\
	&- \sum_{\lambda\kappa}D_{\lambda\kappa, z}^{SS}\bigg[(\boldsymbol{\nabla}_\mu \times \boldsymbol{\nabla}_\lambda)_x( \boldsymbol{\nabla}_\kappa \times \boldsymbol{\nabla}_\nu)_x 
	+(\boldsymbol{\nabla}_\mu \times \boldsymbol{\nabla}_\lambda)_y( \boldsymbol{\nabla}_\kappa \times \boldsymbol{\nabla}_\nu)_y\bigg] (\mu\lambda|\kappa\nu)\nonumber \\
	 &-i \sum_{\lambda\kappa} D_{\lambda\kappa, s}^{SS} \bigg[(\boldsymbol{\nabla}_\mu \cdot \boldsymbol{\nabla}_\lambda)( \boldsymbol{\nabla}_\kappa \times \boldsymbol{\nabla}_\nu)_z +  (\boldsymbol{\nabla}_\mu \times \boldsymbol{\nabla}_\lambda)_z( \boldsymbol{\nabla}_\kappa \cdot \boldsymbol{\nabla}_\nu)\bigg](\mu\lambda|\kappa\nu)\nonumber\\
	&- \sum_{\lambda\kappa} D_{\lambda\kappa, y}^{SS} \bigg[ (\boldsymbol{\nabla}_\mu \cdot \boldsymbol{\nabla}_\lambda)( \boldsymbol{\nabla}_\kappa \times \boldsymbol{\nabla}_\nu)_x
	- (\boldsymbol{\nabla}_\mu \times \boldsymbol{\nabla}_\lambda)_x( \boldsymbol{\nabla}_\kappa \cdot \boldsymbol{\nabla}_\nu)  \bigg] (\mu\lambda|\kappa\nu)\nonumber \\
	&+ \sum_{\lambda\kappa} 
	D_{\lambda\kappa, x}^{SS} \bigg[ (\boldsymbol{\nabla}_\mu \cdot \boldsymbol{\nabla}_\lambda)( \boldsymbol{\nabla}_\kappa \times \boldsymbol{\nabla}_\nu)_y - (\boldsymbol{\nabla}_\mu \times \boldsymbol{\nabla}_\lambda)_y( \boldsymbol{\nabla}_\kappa \cdot \boldsymbol{\nabla}_\nu) \bigg] (\mu\lambda|\kappa\nu)\nonumber \\
	&+i \sum_{\lambda\kappa} D_{\lambda\kappa, s}^{SS} \bigg[(\boldsymbol{\nabla}_\mu \times \boldsymbol{\nabla}_\lambda)_x( \boldsymbol{\nabla}_\kappa \times \boldsymbol{\nabla}_\nu)_y
	- (\boldsymbol{\nabla}_\mu \times \boldsymbol{\nabla}_\lambda)_y( \boldsymbol{\nabla}_\kappa \times \boldsymbol{\nabla}_\nu)_x \bigg](\mu\lambda|\kappa\nu)\nonumber \\
	& + \sum_{\lambda\kappa} D_{\lambda\kappa, x}^{SS} \bigg[(\boldsymbol{\nabla}_\mu \times \boldsymbol{\nabla}_\lambda)_x( \boldsymbol{\nabla}_\kappa \times \boldsymbol{\nabla}_\nu)_z 
	+  (\boldsymbol{\nabla}_\mu \times \boldsymbol{\nabla}_\lambda)_z( \boldsymbol{\nabla}_\kappa \times \boldsymbol{\nabla}_\nu)_x\bigg](\mu\lambda|\kappa\nu) \nonumber \\
	&
	+ \sum_{\lambda\kappa}D_{\lambda\kappa, y}^{SS} \bigg[(\boldsymbol{\nabla}_\mu \times \boldsymbol{\nabla}_\lambda)_y( \boldsymbol{\nabla}_\kappa \times \boldsymbol{\nabla}_\nu)_z 
	+ (\boldsymbol{\nabla}_\mu \times \boldsymbol{\nabla}_\lambda)_z( \boldsymbol{\nabla}_\kappa \times \boldsymbol{\nabla}_\nu)_y  
	\bigg] (\mu\lambda|\kappa\nu)
\end{align}

\begin{align}
    V^{C(4), SS}_{\mu\nu, x} &=
    2\sum_{\lambda\kappa} \bigg[i
	D^{SS}_{\lambda\kappa, s} (\boldsymbol{\nabla}_\mu \times \boldsymbol{\nabla}_\nu)_x (\boldsymbol{\nabla}_\kappa \cdot \boldsymbol{\nabla}_\lambda)
	 -\sum_{J=x,y,z} D_{\lambda\kappa,J}^{SS} (\boldsymbol{\nabla}_\mu \times \boldsymbol{\nabla}_\nu)_x (\boldsymbol{\nabla}_\kappa\times\boldsymbol{\nabla}_\lambda)_J \bigg] (\mu\nu|\kappa\lambda) \nonumber \\
	 & - \sum_{\lambda\kappa} D_{\lambda\kappa, x}^{SS} 
	 \bigg[(\boldsymbol{\nabla}_\mu \cdot \boldsymbol{\nabla}_\lambda)( \boldsymbol{\nabla}_\kappa \cdot \boldsymbol{\nabla}_\nu)-(\boldsymbol{\nabla}_\mu \times \boldsymbol{\nabla}_\lambda)_x( \boldsymbol{\nabla}_\kappa \times \boldsymbol{\nabla}_\nu)_x \bigg](\mu\lambda|\kappa\nu) \nonumber \\
	&-\sum_{\lambda\kappa} D_{\lambda\kappa, x}^{SS}
	\bigg[ (\boldsymbol{\nabla}_\mu \times \boldsymbol{\nabla}_\lambda)_y( \boldsymbol{\nabla}_\kappa \times \boldsymbol{\nabla}_\nu)_y 
	+(\boldsymbol{\nabla}_\mu \times \boldsymbol{\nabla}_\lambda)_z( \boldsymbol{\nabla}_\kappa \times \boldsymbol{\nabla}_\nu)_z \bigg] (\mu\nu|\kappa\lambda) \nonumber \\
	&-i\sum_{\lambda\kappa} D_{\lambda\kappa, s}^{SS}
	\bigg[ 
	(\boldsymbol{\nabla}_\mu \cdot \boldsymbol{\nabla}_\lambda)( \boldsymbol{\nabla}_\kappa \times \boldsymbol{\nabla}_\nu)_x
	+(\boldsymbol{\nabla}_\mu \times \boldsymbol{\nabla}_\lambda)_x( \boldsymbol{\nabla}_\kappa \cdot \boldsymbol{\nabla}_\nu)
	\bigg] (\mu\nu|\kappa\lambda) \nonumber \\
	 &+ \sum_{\lambda\kappa} D_{\lambda\kappa, y}^{SS} 
	 \bigg[ 
	 (\boldsymbol{\nabla}_\mu \cdot \boldsymbol{\nabla}_\lambda)( \boldsymbol{\nabla}_\kappa \times \boldsymbol{\nabla}_\nu)_z 
	 -(\boldsymbol{\nabla}_\mu \times \boldsymbol{\nabla}_\lambda)_z( \boldsymbol{\nabla}_\kappa \cdot \boldsymbol{\nabla}_\nu)
	\bigg] (\mu\nu|\kappa\lambda) \nonumber \\
    &- \sum_{\lambda\kappa} D_{\lambda\kappa, z}^{SS} 
	 \bigg[
    (\boldsymbol{\nabla}_\mu \cdot \boldsymbol{\nabla}_\lambda)( \boldsymbol{\nabla}_\kappa \times \boldsymbol{\nabla}_\nu)_y  
	 -(\boldsymbol{\nabla}_\mu \times \boldsymbol{\nabla}_\lambda)_y( \boldsymbol{\nabla}_\kappa \cdot \boldsymbol{\nabla}_\nu) 
	 \bigg] (\mu\lambda|\kappa\nu) \nonumber \\ 
	&+i\sum_{\lambda\kappa} D_{\lambda\kappa, s}^{SS} 
	 \bigg[
	(\boldsymbol{\nabla}_\mu \times \boldsymbol{\nabla}_\lambda)_y( \boldsymbol{\nabla}_\kappa \times \boldsymbol{\nabla}_\nu)_z 
	-(\boldsymbol{\nabla}_\mu \times \boldsymbol{\nabla}_\lambda)_z( \boldsymbol{\nabla}_\kappa \times \boldsymbol{\nabla}_\nu)_y 
	\bigg] (\mu\lambda|\kappa\nu) \nonumber \\
	&+ \sum_{\lambda\kappa} D_{\lambda\kappa, y}^{SS} 
	 \bigg[
	  (\boldsymbol{\nabla}_\mu \times \boldsymbol{\nabla}_\lambda)_x( \boldsymbol{\nabla}_\kappa \times \boldsymbol{\nabla}_\nu)_y
	+ (\boldsymbol{\nabla}_\mu \times \boldsymbol{\nabla}_\lambda)_y( \boldsymbol{\nabla}_\kappa \times \boldsymbol{\nabla}_\nu)_x 
	\bigg] (\mu\lambda|\kappa\nu) \nonumber \\ 
    &+ \sum_{\lambda\kappa} D_{\lambda\kappa, z}^{SS} 
	 \bigg[
	 (\boldsymbol{\nabla}_\mu \times \boldsymbol{\nabla}_\lambda)_x( \boldsymbol{\nabla}_\kappa \times \boldsymbol{\nabla}_\nu)_z 
	 +(\boldsymbol{\nabla}_\mu \times \boldsymbol{\nabla}_\lambda)_z( \boldsymbol{\nabla}_\kappa \times \boldsymbol{\nabla}_\nu)_x
	 \bigg] (\mu\lambda|\kappa\nu) 
\end{align}

\begin{align}
    V^{C(4), SS}_{\mu\nu, y} &=
    2\sum_{\lambda\kappa}
    \bigg[
    iD^{SS}_{\lambda\kappa, s} (\boldsymbol{\nabla}_\mu \times \boldsymbol{\nabla}_\nu)_y (\boldsymbol{\nabla}_\kappa \cdot \boldsymbol{\nabla}_\lambda)
	 -\sum_{J=x,y,z} D_{\lambda\kappa,J}^{SS} (\boldsymbol{\nabla}_\mu \times \boldsymbol{\nabla}_\nu)_y (\boldsymbol{\nabla}_\kappa\times\boldsymbol{\nabla}_\lambda)_J
	 \bigg] (\mu\nu|\kappa\lambda) \nonumber \\
    &-\sum_{\lambda\kappa}D_{\lambda\kappa, y}^{SS}
    \bigg[
    (\boldsymbol{\nabla}_\mu \cdot \boldsymbol{\nabla}_\lambda)( \boldsymbol{\nabla}_\kappa \cdot \boldsymbol{\nabla}_\nu)
    -(\boldsymbol{\nabla}_\mu \times \boldsymbol{\nabla}_\lambda)_y( \boldsymbol{\nabla}_\kappa \times \boldsymbol{\nabla}_\nu)_y
    \bigg] (\mu\lambda|\kappa\nu) \nonumber \\
    &-\sum_{\lambda\kappa}D_{\lambda\kappa, y}^{SS}
    \bigg[
    (\boldsymbol{\nabla}_\mu \times \boldsymbol{\nabla}_\lambda)_x( \boldsymbol{\nabla}_\kappa \times \boldsymbol{\nabla}_\nu)_x 
	+(\boldsymbol{\nabla}_\mu \times \boldsymbol{\nabla}_\lambda)_z( \boldsymbol{\nabla}_\kappa \times \boldsymbol{\nabla}_\nu)_z
    \bigg] (\mu\lambda|\kappa\nu) \nonumber \\
    &-i\sum_{\lambda\kappa}D_{\lambda\kappa, s}^{SS} 
    \bigg[
    (\boldsymbol{\nabla}_\mu \cdot \boldsymbol{\nabla}_\lambda)( \boldsymbol{\nabla}_\kappa \times \boldsymbol{\nabla}_\nu)_y
    + (\boldsymbol{\nabla}_\mu \times \boldsymbol{\nabla}_\lambda)_y( \boldsymbol{\nabla}_\kappa \cdot \boldsymbol{\nabla}_\nu)
    \bigg] (\mu\lambda|\kappa\nu) \nonumber \\
    &-\sum_{\lambda\kappa}D_{\lambda\kappa, x}^{SS}
    \bigg[
    (\boldsymbol{\nabla}_\mu \cdot \boldsymbol{\nabla}_\lambda)( \boldsymbol{\nabla}_\kappa \times \boldsymbol{\nabla}_\nu)_z
    -(\boldsymbol{\nabla}_\mu \times \boldsymbol{\nabla}_\lambda)_z( \boldsymbol{\nabla}_\kappa \cdot \boldsymbol{\nabla}_\nu)
    \bigg] (\mu\lambda|\kappa\nu) \nonumber \\
    &+\sum_{\lambda\kappa}D_{\lambda\kappa, z}^{SS}
    \bigg[
    (\boldsymbol{\nabla}_\mu \cdot \boldsymbol{\nabla}_\lambda)( \boldsymbol{\nabla}_\kappa \times \boldsymbol{\nabla}_\nu)_x
    -(\boldsymbol{\nabla}_\mu \times \boldsymbol{\nabla}_\lambda)_x( \boldsymbol{\nabla}_\kappa \cdot \boldsymbol{\nabla}_\nu)
	\bigg] (\mu\lambda|\kappa\nu) \nonumber \\
	& -i \sum_{\lambda\kappa}D_{\lambda\kappa, s}^{SS}
    \bigg[
    (\boldsymbol{\nabla}_\mu \times \boldsymbol{\nabla}_\lambda)_x( \boldsymbol{\nabla}_\kappa \times \boldsymbol{\nabla}_\nu)_z 
	- (\boldsymbol{\nabla}_\mu \times \boldsymbol{\nabla}_\lambda)_z( \boldsymbol{\nabla}_\kappa \times \boldsymbol{\nabla}_\nu)_x 
	\bigg] (\mu\lambda|\kappa\nu) \nonumber \\
	&+ \sum_{\lambda\kappa}D_{\lambda\kappa, x}^{SS}
    \bigg[
    (\boldsymbol{\nabla}_\mu \times \boldsymbol{\nabla}_\lambda)_x( \boldsymbol{\nabla}_\kappa \times \boldsymbol{\nabla}_\nu)_y
	+ (\boldsymbol{\nabla}_\mu \times \boldsymbol{\nabla}_\lambda)_y( \boldsymbol{\nabla}_\kappa \times \boldsymbol{\nabla}_\nu)_x 
	\bigg] (\mu\lambda|\kappa\nu) \nonumber \\
	&+ \sum_{\lambda\kappa}D_{\lambda\kappa, z}^{SS}
    \bigg[
	(\boldsymbol{\nabla}_\mu \times \boldsymbol{\nabla}_\lambda)_y( \boldsymbol{\nabla}_\kappa \times \boldsymbol{\nabla}_\nu)_z 
	+ (\boldsymbol{\nabla}_\mu \times \boldsymbol{\nabla}_\lambda)_z( \boldsymbol{\nabla}_\kappa \times \boldsymbol{\nabla}_\nu)_y 
	\bigg] (\mu\lambda|\kappa\nu)\
\end{align}

\subsection*{Appendix C: Cost Analysis of Dirac-Hartree--Fock in Unrestricted-Kinetic-Balanced Scalar Basis}

The computational cost of the four-component method using the unrestricted-kinetic-balanced (UKB) scalar basis strongly depends on the nature of the implementation. This is because upon the generation of small component basis, the kinetic-balance condition no longer directly appears in the two-spinor basis. This allows an implementation in scalar basis to adapt various techniques developed for non-relativistic methods. This is especially true for the Dirac-Coulomb term where Pauli spinors do not enter the integrals.

In this analysis, we assume that all integrals and integral-density contractions in UKB scalar basis are in the non-relativistic one-component framework, and all integrals are stored as real-valued quantities. 
In this discussion, $N_L$ and $N_S$ are the numbers of basis functions for the large and small component, respectively, in UKB scalar basis. We also assume that both the Fock and density matrices are in the quaternion representation so that the following expressions represent the optimal algorithm in terms of both memory requirement and FLOP count.

For the Coulomb operator, the contraction pattern can be separated into Coulomb $\mathbf{J}^C$ and exchange $\mathbf{K}^C$ matrices,
\begin{align}
J_{\mu\nu}^{C,LL} &= \sum_{\gamma=0,1,2,3} \sum_{\kappa \lambda} (\mu^L \nu^L|\kappa^L \lambda^L) D^{LL}_{\lambda\kappa, \gamma}\hat{\mathbf{e}}_\gamma
                + \sum_{\gamma=0,1,2,3} \sum_{\kappa \lambda} (\mu^L \nu^L|\kappa^S \lambda^S) D^{SS}_{\lambda\kappa, \gamma}\hat{\mathbf{e}}_\gamma
\\
J_{\mu\nu}^{C,SS} &= \sum_{\gamma=0,1,2,3} \sum_{\kappa \lambda} (\mu^S \nu^S|\kappa^L \lambda^L) D^{LL}_{\lambda\kappa, \gamma}\hat{\mathbf{e}}_\gamma
                + \sum_{\gamma=0,1,2,3} \sum_{\kappa \lambda} (\mu^S \nu^S|\kappa^S \lambda^S) D^{SS}_{\lambda\kappa, \gamma}\hat{\mathbf{e}}_\gamma\\
K_{\mu\nu}^{C,LL} &= \sum_{\gamma=0,1,2,3} \sum_{\kappa \lambda} (\mu^L \lambda^L|\kappa^L \nu^L) D^{LL}_{\lambda\kappa, \gamma}\hat{\mathbf{e}}_\gamma
\\
K_{\mu\nu}^{C,LS} &= \sum_{\gamma=0,1,2,3} \sum_{\kappa \lambda} (\mu^L \lambda^L|\kappa^S \nu^S) D^{LS}_{\lambda\kappa, \gamma}\hat{\mathbf{e}}_\gamma
\\
K_{\mu\nu}^{C,SL} &= \sum_{\gamma=0,1,2,3} \sum_{\kappa \lambda} (\mu^S \lambda^S|\kappa^L \nu^L) D^{SL}_{\lambda\kappa, \gamma}\hat{\mathbf{e}}_\gamma
\\
K_{\mu\nu}^{C,SS} &= \sum_{\gamma=0,1,2,3} \sum_{\kappa \lambda} (\mu^S \lambda^S|\kappa^S \nu^S) D^{SS}_{\lambda\kappa, \gamma}\hat{\mathbf{e}}_\gamma
\end{align}
where $(\hat{\mathbf{e}}_0, \hat{\mathbf{e}}_1, \hat{\mathbf{e}}_2, \hat{\mathbf{e}}_3)$ is the quaternion basis of choice. Based on these expressions, constructing the Coulomb $\mathbf{J}^C$ matrix requires $4\times(N_L^4 + 2N_L^2N_S^2 + N_S^4)$ number of 
integral-density contraction operations, and the $\mathbf{K}$ matrix requires $4\times(N_L^4+2N_L^2N_S^2+N_S^4)$ operations, where the factor of 4 comes from the contraction with four quaternion components. The memory requirement for building the Dirac-Coulomb matrix is $N_L^4+N_L^2N_S^2+N_S^4$. All $N_S^4$ quantities in the analysis correspond to the contributions from the $(SS|SS)$ term.

In order to build the Gaunt operator in UKB scalar basis with the lowest computational cost, the Pauli spin matrices are also cast in the quaternion representation,
\begin{equation*}
\boldsymbol{\sigma} = \begin{pmatrix}\sum_\gamma \sigma_{x,\gamma} \mathbf{\hat{e}}_\gamma\\
\sum_\gamma \sigma_{y,\gamma} \mathbf{\hat{e}}_\gamma\\
\sum_\gamma \sigma_{z,\gamma} \mathbf{\hat{e}}_\gamma\end{pmatrix}.
\end{equation*}
As a result, the Gaunt operator under UKB scalar basis can be evaluated with the
real scalar ERIs, 
\begin{align}
J_{\mu\nu,\tau}^{G,LS} &= \sum_{\gamma=0,1,2,3} \mathbf{\hat{e}}_\tau\circ \boldsymbol\sigma(1)\cdot\big[\boldsymbol{\sigma}(2)\circ (\mu^L \nu^S|\kappa^L \lambda^S) (D^{LS}_{\kappa\lambda,\gamma}\hat{\mathbf{e}}_\gamma + D^{SL}_{\lambda\kappa,\gamma}\hat{\mathbf{e}}_\gamma)\big]
\\
K_{\mu\nu,\tau}^{G,LL} &= \sum_{\gamma=0,1,2,3} \mathbf{\hat{e}}_\tau\circ \boldsymbol\sigma(1)\cdot\big[\boldsymbol{\sigma}(2)\circ  (\mu^L \lambda^S | \nu^L \kappa^S) D^{SS}_{\lambda\kappa,\gamma}\hat{\mathbf{e}}_\gamma\big] 
\\
K_{\mu\nu,\tau}^{G,SS} &= \sum_{\gamma=0,1,2,3} \mathbf{\hat{e}}_\tau\circ \boldsymbol\sigma(1)\cdot\big[\boldsymbol{\sigma}(2)\circ (\lambda^L \mu^S | \kappa^L \nu^S) D^{LL}_{\lambda\kappa,\gamma}\hat{\mathbf{e}}_\gamma\big] 
\\
K_{\mu\nu,\tau}^{G,LS} &= \sum_{\gamma=0,1,2,3} \mathbf{\hat{e}}_\tau\circ \boldsymbol\sigma(1)\cdot\big[\boldsymbol{\sigma}(2)\circ  (\mu^L \lambda^S | \kappa^L \nu^S) D^{SL}_{\lambda\kappa,\gamma}\hat{\mathbf{e}}_\gamma\big] 
\\
K_{\mu\nu,\tau}^{G,SL} &= \sum_{\gamma=0,1,2,3} \mathbf{\hat{e}}_\tau\circ \boldsymbol\sigma(1)\cdot\big[\boldsymbol{\sigma}(2)\circ (\lambda^L \mu^S | \nu^L \kappa^S) D^{LS}_{\lambda\kappa,\gamma}\hat{\mathbf{e}}_\gamma\big]
\end{align}
where we use `$\circ$' to represent  dot-product in the quaternion space and `$\cdot$' for dot-product in the Pauli spinor space. It is clear that building the $\mathbf{J}^G$ and $\mathbf{K}^G$ matrix requires $4N_L^2N_S^2$ and $16 N_L^2N_S^2$ numbers of integral-density contraction operations with a memory requirement of only $N_L^2N_S^2$.

\begin{table}[ht]
{
\caption{Calculated Dirac-Hartree-Fock ground state energies (Hartrees) for the hydrogen atom and a series of hydrogenic cations compared with the analytical solution of the Dirac equation. The uncontracted Sapporo-DKH3-QZP-2012-no basis is used for all elements except actinides (Ac$^{88+}$, Es$^{98+}$, Md$^{100+}$). Actinide calculations use  uncontracted cc-pwCVDZ-X2C basis\cite{Peterson17_084108}. Seven steep $s$ functions are added to the uncontracted basis. Point charge nuclear model is used.} 
\centering 
\begin{tabular}{l r r r} 
\hline
element & \adjustbox{stack=ll}{analytical\\solution} & \adjustbox{stack=ll}{uncontracted\\basis} & \adjustbox{stack=ll}{percent error\\($\times 10^{-3}\%$)}  \\ [1.0ex] 
\hline 
H &   $   -0.5000067$  & $   -0.5000029$ & $0.75$  \\
B$^{4+}$ &   $  -12.5041630$ & $  -12.5038524$ & $2.48$ \\
F$^{8+}$ &   $  -40.5437672$ & $  -40.5428584$ & $2.24$ \\
Al$^{12+}$ &  $  -84.6909743$ & $  -84.6891918$ & $2.10$ \\
Cl$^{16+}$ &  $ -145.0602703$ & $ -145.0573482$ & $2.01$ \\
Mn$^{24+}$ &  $ -315.1443548$ & $ -315.1385013$ & $1.86$ \\
Rb$^{26+}$ &  $ -697.4517670$ & $ -697.4334697$ & $2.62$ \\
Mo$^{41+}$ &  $ -903.7467394$ & $ -903.7213445$ & $2.81$\\
In$^{48+}$ &  $-1241.5414919$ & $-1241.4977248$ & $3.52$ \\
Cs$^{54+}$ &  $-1578.8736026$ & $-1578.8504245$ & $1.47$ \\
Ta$^{72+}$ &  $-2886.3133130$ & $-2886.2556854$ & $2.00$ \\
Ac$^{88+}$ &  $-4499.5659601$ & $-4499.5624725$ & $0.08$ \\
Es$^{98+}$ &  $-5794.4853497$ & $-5794.4685369$ & $0.29$  \\
Md$^{100+}$ &  $-6087.0342057$ & $-6087.0107482$ & $0.39$
\\[1ex] 
\hline 
\end{tabular}

\label{table:hydrogenicsteep} 
}
\end{table}

\subsection*{Appendix D: Calculations of Hydrogenic Cations/Atom Compared with Analytical Results of Dirac Equation}

Energy levels of the relativistic Dirac hydrogen atom and hydrogenic cations can be evaluated analytically using the following relation (in atomic units)
\begin{equation}
    E = c^2 \times \bigg[ \frac{1}{\sqrt{1+(\frac{Z \alpha}{n-(j+1/2)+\sqrt{(j+1/2)^2-Z^2\alpha^2}})^2}}   -1 \bigg]
\end{equation}
where $Z$ is the nuclear charge and $\alpha$ is fine structure constant. For one electron systems in their ground states, $n=1$ and $j=\frac{1}{2}$.  In order to validate the four-component Dirac--Hartree--Fock implementation, we compare the computed energies with analytical values in \cref{table:hydrogenicsteep}. For a better description of the steep electronic wave function near the nucleus, seven additional steep $s$ functions are added to the uncontracted basis set, with exponents ranging from 3$^1$ to $3^7$ times the largest exponent in the original basis.\cite{Kaupp08_104101,Liu09_244113} From  this  series,  we  see  that  the  largest  error  between  the  analytical  result  and  the Dirac Hartree-Fock result is 0.00352 \% for In$^{48+}$ when a uncontracted basis is used, showing good agreement and validating our implementation.

\subsection*{Appendix E: Basis Set Limit}

In order to compare our results to the ``true'' basis set limit, we can look at the results from Visscher\cite{Dyall97_207} and Hess\cite{Hess94_183}  that utilized a numerical grid basis for both point nuclei and finite Gaussian nuclei.  As shown in \cref{table:au_atom_data}, we can see that using an uncontracted basis set gives the expected variational upper bound to the numerically exact limit. The error in the ground state energy is only $\sim$0.015 $E_h$ for the Gaussian nuclei model, with all three representations (RKB Pauli, RKB spinor, UKB scalar) behaving similarly. (Additionally the UKB scalar calculation includes the average of configuration method for open shell systems.)


\begin{table}[htbp]
{
{\footnotesize
\begin{center}
\caption{Ground state energies (Hartrees) for Au atom with the Dirac-Coulomb Hamiltonian with a Gaussian nuclear model uncontracted ANO-RCC basis is used. RKB Pauli, RKB spinor, and UKB scalar results were obtained using ChronusQ,\cite{Li20_e1436} PySCF,\cite{Chan18_e1340} and DIRAC,\cite{DIRAC19} respectively. Note that the UKB scalar basis is transformed to the RKB spinor during the orthonomalization step.\cite{DIRAC20}} 
\begin{tabular}{l l  l} 
\hline
Model & Energy \\ [0.5ex] 
\hline 
Numerical Grid [Ref. \citenum{Dyall97_207}] &   $-$19035.59510\\
RKB Pauli   &  $-$19035.57905  \\
RKB Spinor  &   $-$19035.57905  \\
UKB Scalar  &   $-$19035.57807\\
[1ex] 
\hline 
\end{tabular}
{\footnotesize
}
\label{table:au_atom_data}
\end{center}
}
}
\end{table}

\clearpage 
\pagebreak
\bibliography{Journal_Short_Name,Li_Group_References,Sun,Petrone_References} 

\begin{tocentry}
\vspace{-4pt}
\centering
\includegraphics[width=2.75in]{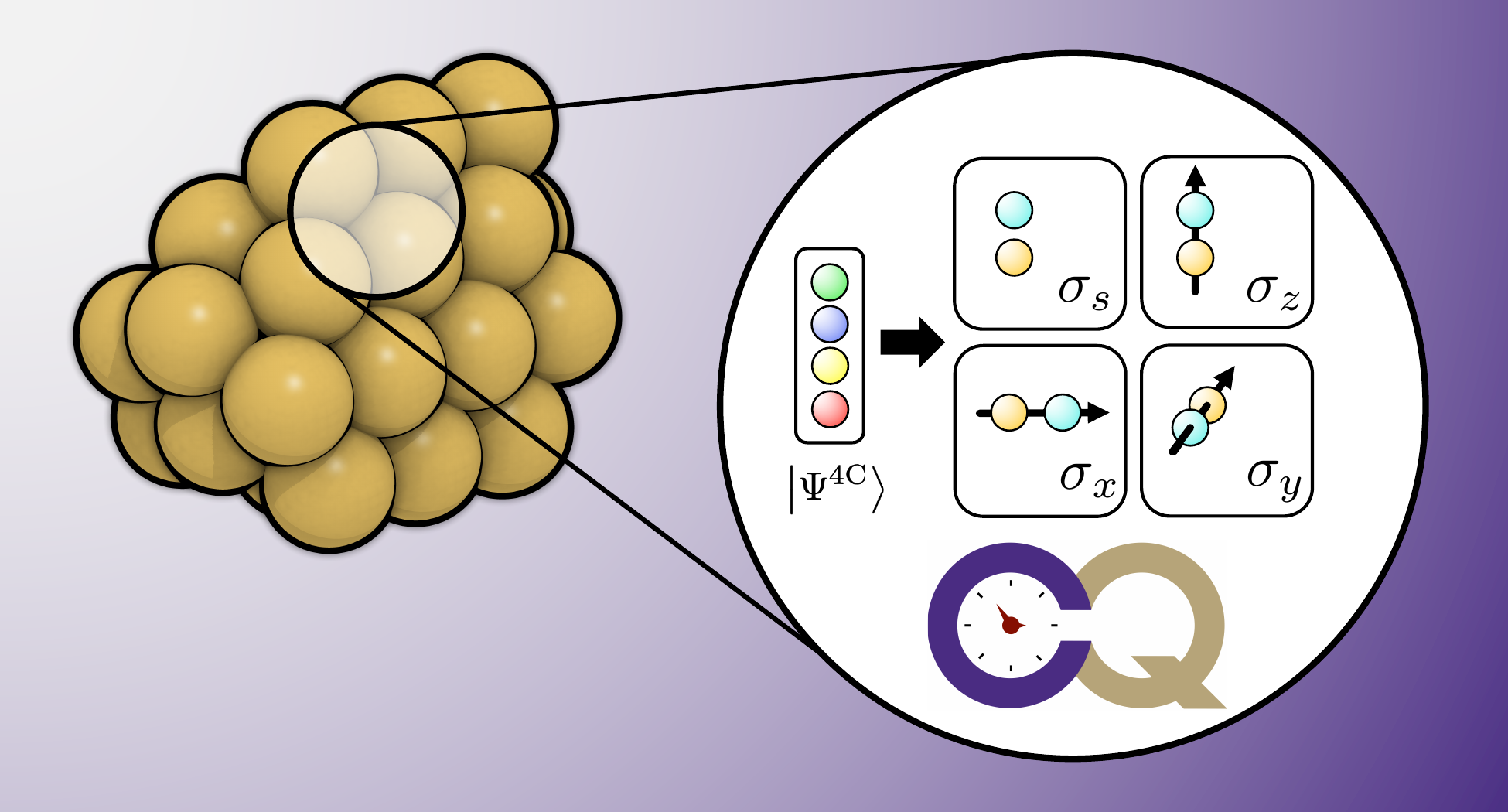}

\end{tocentry}


\end{document}